\documentclass[10pt,a4paper]{article}%
\usepackage[english]{babel}%
\usepackage{amssymb,amsmath,ae,amsthm}%
\usepackage[dvips]{graphicx,color}%
\newcommand{\texorpdfstring}[2]{#1}
\newcommand{\figdraft}{false}%
\newcommand{\figfile}[1]{#1}%
\theoremstyle{plain}%
\newtheorem{theorem}{Theorem}[section]%
\newtheorem{corollary}[theorem]{Corollary}%
\newtheorem{observation}[theorem]{Observation}%
\newtheorem{lemma}[theorem]{Lemma}%
\newtheorem{definition}[theorem]{Definition}%
\newtheorem{assumption}[theorem]{Assumption}%
\newtheorem{remark}[theorem]{Remark}%
\newtheorem{example}[theorem]{Example}%
\definecolor{colorGreen}{rgb}{0.,0.67,0}
\definecolor{colorRed}{rgb}{0.67,0.,0}
\definecolor{colorBlue}{rgb}{0.,0.,0.67}

\newcommand{\toda}{{\rm Toda}}
\newcommand{\crit}{{\rm crit}}

\newcommand{\harm}{{\rm harm}}

\newcommand{\nl}{{\rm nl}}

\newcommand{\tdots}{{...}}%

\newcommand{\CL}{\mathrm{CL}}
\newcommand{\bgr}{{\mathrm{bg}}}
\newcommand{\av}{\mathrm{av}}

\newcommand{\pair}[2]{{\left({#1},\,{#2}\right)}}
\newcommand{\skp}[2]{{\left\langle{#1},\,{#2}\right\rangle}}

\newcommand{\bpair}[2]{{\big({#1},\,{#2}\big)}}
\newcommand{\at}[1]{{\left({#1}\right)}}
\newcommand{\nat}[1]{(#1)}
\newcommand{\bat}[1]{{\big(#1\big)}}
\newcommand{\Bat}[1]{{\Big(#1\Big)}}

\newcommand{\ato}[1]{{\left[{#1}\right]}}

\newcommand{\abs}[1]{\left|{#1}\right|}
\newcommand{\babs}[1]{\big|{#1}\big|}
\newcommand{\nabs}[1]{|{#1}|}

\newcommand{\specA}{\Theta}

\newcommand{\Len}{L}

\newcommand{\iu}{\mathtt{i}}
\newcommand{\mhexp}[1]{{{\mathtt{e}}^{#1}}}

\newcommand{\deq}{:=}
\newcommand{\phase}{{\varphi}}

\newcommand{\ol}[1]{{\overline{#1}}}

\newcommand{\widebar}[1]{{\bar{#1}}}

\newcommand{\norm}[1]{\parallel\!{#1}\!\parallel}

\newcommand{\dint}[1]{\,\mathrm{d}#1}

\newcommand{\bigpar}{\par\quad\par}

\newcommand{\fspace}[1]{{\mathsf{#1}}}
\newcommand{\fspaceL}{\fspace{L}}
\newcommand{\fspaceC}{\fspace{C}}
\newcommand{\fspaceW}{\fspace{W}}

\newcommand{\Rset}{{\mathbb{R}}}

\newcommand{\Nset}{{\mathbb{N}}}

\newcommand{\ocinterval}[2]{(#1,\,#2]}%
\newcommand{\cointerval}[2]{[#1,\,#2)}%
\newcommand{\ccinterval}[2]{[#1,\,#2]}%

\newcommand{\DO}[1]{{O\at{#1}}}
\newcommand{\Do}[1]{{o\at{#1}}}
\newcommand{\nDO}[1]{{O\nat{#1}}}
\newcommand{\nDo}[1]{{o\nat{#1}}}

\newcommand{\al}{{\alpha}}
\newcommand{\be}{{\beta}}
\newcommand{\ga}{{\gamma}}

\newcommand{\eps}{{\varepsilon}}

\newcommand{\la}{{\lambda}}
\newcommand{\si}{{\sigma}}

\newcommand{\om}{{\omega}}

\newcommand{\calA}{\mathcal{A}}
\newcommand{\calB}{\mathcal{B}}

\newcommand{\calK}{\mathcal{K}}
\newcommand{\calL}{\mathcal{L}}
\newcommand{\calM}{\mathcal{M}}
\newcommand{\calN}{\mathcal{N}}

\newcommand{\calP}{\mathcal{P}}

\newcommand{\calS}{\mathcal{S}}
\newcommand{\calT}{\mathcal{T}}
\newcommand{\calU}{\mathcal{U}}

\newcommand{\calW}{\mathcal{W}}

\usepackage{float}%
\begin{document}
%
%
% ------------------------------------------------------------------
%                      Header
% ------------------------------------------------------------------
%
%
%
\title{%
    Unimodal wave trains and solitons in convex FPU chains
}%
\author{Michael Herrmann\thanks{ %
    University of Oxford,
    Mathematical Institute, Centre for Nonlinear PDE (OxPDE),
    24-29 St Giles', Oxford OX1 3LB, United Kingdom, michael.herrmann@maths.ox.ac.uk.}
}%
\maketitle
%
%
%
% ------------------------------------------------------------------
%                      abstract
% ------------------------------------------------------------------
%
%
\begin{abstract}
We consider atomic chains with nearest neighbour interactions and
study periodic  and homoclinic travelling waves which are called
wave trains and solitons, respectively. Our main result is a new
existence proof which relies on the constrained maximisation of the
potential energy and exploits the invariance properties of an
improvement operator. The approach is restricted to convex
interaction potentials but refines the standard results as it
provides the existence of travelling waves with unimodal and even
profile functions. Moreover, we discuss the numerical approximation
and complete localization of wave trains, and show that wave trains
converge to solitons when the periodicity length tends to infinity.
\end{abstract}
%
%
% ------------------------------------------------------------------
%                      MSC and keywords
% ------------------------------------------------------------------
%
\quad\newline\noindent%
\begin{minipage}[t]{0.15\textwidth}%
{\small Keywords:} %
\end{minipage}%
\begin{minipage}[t]{0.8\textwidth}%
\small %
\emph{Fermi-Pasta-Ulam chain}, %
\emph{lattice travelling waves},\\ %
\emph{constrained optimisation}, %
\emph{complete localisation} %
\end{minipage}%
\medskip
\newline\noindent
\begin{minipage}[t]{0.15\textwidth}%
\small MSC (2000): %
\end{minipage}%
\begin{minipage}[t]{0.8\textwidth}%
\small %
37K60, %Lattice dynamics (in 37Kxx=Infinite-dimensional Hamiltonian systems)
47J30, %Variational methods (in 47Jxx=equations and inequalities involving nonlinear operators)
70F45, %Infinite particle systems  (in 70-xx=Mechanics of particles and systems)
74J30 %Nonlinear waves (in 74-xx=mechanics of deformable solids)
\end{minipage}%
%
%
%
% ------------------------------------------------------------------
%                      table of contents
% ------------------------------------------------------------------
%
%
\setcounter{tocdepth}{5} %
\setcounter{secnumdepth}{4}
{\scriptsize{\tableofcontents}}%
%
%
% ------------------------------------------------------------------
%                      contents
% ------------------------------------------------------------------
%
%
%
%
%--------------------------------------------------------------------
\section{Introduction}
%--------------------------------------------------------------------
%
%
We consider infinite chains of identical atoms with unit mass that
are coupled by nearest neighbour interactions. The dynamics in such
chains is governed by Newton's equations
\begin{align}
\label{Intro.FPU1}
\ddot{x}_j\at{t}=\Phi^\prime\bat{{x}_{j+1}\at{t}-{x}_j\at{t}}-
\Phi^\prime\bat{{x}_{j}\at{t}-{x}_{j-1}\at{t}},
\end{align}
where $x_j\at{t}$ denotes the position of the $j$th atom at time
$t$, and $\Phi$ is the interaction potential. Restating
\eqref{Intro.FPU1} in terms of atomic distances
$r_j\at{t}={x}_{j+1}\at{t}-{x}_j\at{t}$ and atomic velocities
$v_j\at{t}=\dot{x}_j\at{t}$ we find
\begin{align}
\label{Intro.FPU2} %
\dot{r}_j\at{t}={v}_{j+1}\at{t}-{v}_j\at{t},
\qquad
\dot{v}_j\at{t}=
\Phi^\prime\bat{{r}_{j}\at{t}}-\Phi^\prime\bat{{r}_{j-1}\at{t}}.
\end{align}
In this article we allow for arbitrary convex interaction potentials
$\Phi$ and refer to \eqref{Intro.FPU1} as Fermi-Pasta-Ulam (FPU)
chain although the potential in the original paper \cite{FPU55} was
a quartic polynomial.
\bigpar
FPU chains can be viewed as simple toy models for crystals and
solids and allow to study some essential properties of
nonlinear elastic materials. Even though \eqref{Intro.FPU1} is a
strong simplification of a real material it obeys a very complex
behaviour and currently we are far from a complete understanding of
its dynamical properties.
\par
During the last decades, a lot of research addressed the existence
and properties of travelling waves in FPU chains because they can be
viewed as elementary waves and provide a lot of insight into the
energy transport in nonlinear media. Such travelling waves solve
nonlinear advance-delay differential equations and it is a
fundamental mathematical problem to characterise the solution set to
those equations.
\par
\begin{figure}[ht!]
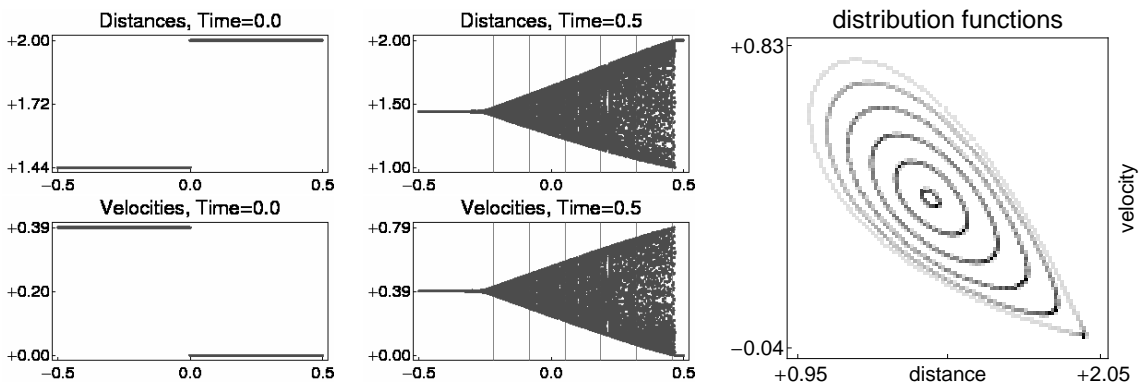
%
\centering{%
\begin{minipage}[c]{0.275\textwidth}%
\includegraphics[width=\textwidth, draft=\figdraft]%
{\figfile{dist_mi_0}}%
\\%
\includegraphics[width=\textwidth, draft=\figdraft]%
{\figfile{vel_mi_0}}%
\end{minipage}%
\hspace{0.025\textwidth}%
\begin{minipage}[c]{0.275\textwidth}%
\includegraphics[width=\textwidth, draft=\figdraft]%
{\figfile{dist_mi_1}}%
\\%
\includegraphics[width=\textwidth, draft=\figdraft]%
{\figfile{vel_mi_1}}%
\end{minipage}%
\hspace{0.025\textwidth}%
\begin{minipage}[c]{0.345\textwidth}%
\includegraphics[width=\textwidth, draft=\figdraft]%
{\figfile{df_dist_vel}}%
\end{minipage}%
}%
\caption{The atoms in a dispersive shock self-organize into
modulated travelling waves. \emph{Left}. Snapshots of atomic
distances and velocities against the scaled particle index for
different macroscopic times. \emph{Right}. Corresponding family of
wave trains with super-sonic soliton as `homoclinic limit'; the
picture shows the density plots of six local distribution functions
in the $\pair{r}{v}$-plane. Modulation theory predicts that each of
these local distribution functions corresponds to
a travelling wave whose parameters depend on macroscopic time and particle index.}%
\label{Fig:DispersiveShock}
\end{figure}

A further motivation for the study of travelling waves is related to
atomistic Riemann problems and self-thermalisation of FPU chains:
Starting with piecewise constant initial data for the atomic
distances and velocities, solutions to \eqref{Intro.FPU2} are
self-similar on a macroscopic scale and involve \emph{dispersive
shocks}, that are fan-like structures with strong microscopic
oscillations, see Figure \ref{Fig:DispersiveShock}. It is known from
the theory of integrable systems and numerical simulations that the
oscillations within a dispersive shock can be described by
\emph{modulated travelling waves}, compare \cite{FV99,DH08} and
references therein.
%
%
%
%--------------------------------------------------------------------
\paragraph*{Travelling waves and main result}
%--------------------------------------------------------------------
%
%
%
\emph{Travelling waves} are exact solutions to \eqref{Intro.FPU2}
and satisfy the ansatz\footnote{The ansatz \eqref{Intro:TWAnsatz} is
slightly more general than the usual one, which assumes that the
atomic positions $x_j$ depend on the phase variable $\phase$. In
fact, the ansatz $x_j\at{t}={X}\at{k{j}+\om{t}}$ is \emph{not}
invariant under Galilean transformations
$x\rightsquigarrow{x}+{v_0}t$, whereas \eqref{Intro:TWAnsatz} has
this property as it respects
$r\rightsquigarrow{r},v\rightsquigarrow{v}+{v_0}$. }
\begin{align}
\label{Intro:TWAnsatz} %
r_j\at{t}={R}\at\phase,
\qquad%
v_j\at{t}={V}\at\phase,
\qquad%
\phase={k}j+\om{t},
\end{align}
with phase $\phase$, wave number $k$, (negative) frequency $\om$,
and profile functions $R$ and $V$. Inserting \eqref{Intro:TWAnsatz}
into \eqref{Intro.FPU2} we obtain the nonlinear advance-delay
differential equations
\begin{align}
\label{Intro:TWEqn}
\om\frac{\dint}{\dint\phase}{R}=\nabla_k^+{V},\qquad
\om\frac{\dint}{\dint\phase}{V}=\nabla_k^-\Phi^\prime\at{{R}},
\end{align}
where $\nabla_k^+$ and $\nabla_k^-$ denote the forward and backward
difference operators with shift $k$, respectively. Depending on the
properties of the profile functions we distinguish the following
cases:
\begin{enumerate}
\item
\emph{Wave trains} or \emph{periodic waves}: %
${R}$ and ${V}$ are periodic,
\item
\emph{Solitons} or \emph{homoclinic waves}: %
${R}$ and ${V}$ are localized over a constant background state,
\item
\emph{Fronts} or \emph{heteroclinic waves}: %
${R}$ and ${V}$ connect different constant background states,
\item
\emph{Oscillatory fronts}: ${R}$ and ${V}$ connect different
asymptotic wave trains.
\end{enumerate}
Notice that our usage of 'soliton' is quite sloppy: Localized
travelling waves are sometimes called `solitary waves', and `soliton'
then refers to a solitary waves that survives collisions with other
such waves unchanged.
\bigpar
In this article we show the existence of wave trains and solitons
with unimodal and even profile functions $R$ and $V$, where
\emph{even} means as usual invariance under
$\phase\rightsquigarrow-\phase$, and \emph{unimodal} functions are
monotone for both $\phase\leq{0}$ and $\phase\geq{0}$. Our main
result can be stated as follows.
\begin{theorem}
\label{Intro:MainResult} Under natural regularity assumptions on the
convex potential $\Phi$ there exists a four-parameter family of wave
trains, and if $\Phi$ satisfies additionally some super-quadratic
grow conditions then there exists also a three-parameter family of
solitons. Moreover, the profile functions $R$ and $V$ for both
families are unimodal and even.
\end{theorem}
Close to our work are \cite{FW94,SW97,FV99,PP00,SZ07} where the
existence of travelling waves is likewise studied in a variational
framework. Therefore we shall compare both our method
(\S\ref{sec:overview.variational}) and results
(\S\ref{sec:ES:GSQCrit}) with those presented therein. We also refer
to the numerical study \cite{DEFW93}, to \cite{FM03} for existence
results in 2D lattices, and to \cite{Ioo00} which proves the
existence of small-amplitude travelling waves by means of
center-manifold reduction. Moreover, the existence of fronts is
studied in \cite{HR08b}, and \cite{SZ08} concerns oscillatory fronts
in FPU chains with bi-harmonic potentials.
\par
All results presented below concern solely wave trains and solitons
in FPU chains, but the method can also be applied to other
Hamiltonian lattices with convex potential energy $\calP$ as for
instance Klein-Gordon chains with convex on-site potential, see
\cite{Her09b}, and atomic chains with next-nearest neighbour
interactions.
\par
We emphasize that we are not able to provide uniqueness results
for travelling waves. Uniqueness of relative equilibria in
Hamiltonian lattices is a notoriously difficult problem, and almost
nothing is known about it. The only available results concern either
the near-sonic limit \cite{FP99}, or systems where the travelling
wave equation can be solved explicitly. Examples are the Toda chain
\cite{Tod81}, the discrete nonlinear Schr\"odinger equation
\cite{HLM94}, the harmonic chain, and the hard-sphere model
\cite{DHM06}.
%
%
%--------------------------------------------------------------------
\paragraph*{Overview on the proof and organisation of the paper}
%--------------------------------------------------------------------
%
%
%
In a first step we reformulate the
travelling wave equation \eqref{Intro:TWEqn} in terms of a
normalised profile function $W\in\fspaceL^2$. More precisely, 
we show that \eqref{Intro:TWEqn} can be
transformed into a nonlinear eigenvalue equation
\begin{align}
\label{Intro.EigenValueProblem}
\om^2{W}=\partial\calP\at{W},
\end{align}
where $\calP\at{W}$ is the potential energy of a travelling wave.
The profile function $W$ has no physical meaning, but determines $R$
and $V$ via
$R\at\phase=r_0+\int_\phase^{\phase+k}{W}\at{\tilde\phase}\dint{\tilde\phase}$
and $V\at\phase=v_0+\om{W}\at\phase$, where $r_0$ and $v_0$ are
suitable normalization constants.
\bigpar
Our approach relies on a combination of
variational and dynamical concepts, and can be summarized
as follows.
\begin{enumerate}
\item
Equation \eqref{Intro.EigenValueProblem} is the Euler-Lagrange
equation for the optimisation problem $\calP\at{W}\to\max$ subjected
to the constraint
\begin{math}
W\in\calB_\ga, %
\end{math} %
where $\ga$ is a free parameter, $\calB_\ga\subset\fspaceL^2$
denotes the ball of radius $\sqrt{2\ga}$ , and $\om^2$ is the
Lagrangian multiplier.
\item
There exists an improvement dynamics
\begin{math}
W\mapsto\calT_\ga\ato{W}
\end{math}
on $\calB_\ga$ that increases the potential energy. Moreover, each
stationary point of this dynamics solves
\eqref{Intro.EigenValueProblem}, and vise versa.
\item
There exist non-trivial cones $S$ that are invariant under the
improvement dynamics. Consequently, each maximiser of $\calP$ in
$\calS\cap\calB_\ga$ is a travelling wave, see Theorem
\ref{IS:Cone2.Theo}.
\end{enumerate}
We emphasize that the convexity of $\Phi$ is essential for this approach as it is
intimately related to both the properties of $\calT_\ga$ and the
existence of non-trivial invariant cones.
\bigpar
A major part of the mathematical analysis done in this paper is
needed to show that there exists a maximizer of $\calP$ in
$\calS_\ga$. In the wave-train setting we can use rather simple
compactness arguments as the functional $\calP$ is continuous with
respect to the weak topology in $\fspaceL^2$. In the soliton
setting, however, we lack the weak compactness of $\calP$, and the
existence proof for maximizers requires more sophisticated
arguments. Our main technical result in this context is Lemma
\ref{ES:Tightness.Lemma1}, which implies that (for certain $S$) the
maximising sequences for $\calP$ in $\calS_\ga$ are
\emph{localized}, and hence pre-compact in the strong topology.
\bigpar%
This paper is organised as follows. In \S\ref{sec:overview.reform} we derive
the fixed point equation \eqref{Intro.EigenValueProblem} for both wave trains and solitons.
The details of our variational approach are presented in
\S\ref{sec:overview.variational}, and \S\ref{sec:tools} concerns the analytical properties 
of the underlying functionals and operators.
In \S\ref{sec:WT.existence} we continue with the
existence proof for wave trains and 
present some numerical simulations; 
the \emph{complete localisation} of wave
trains is studied in \S\ref{sec:WT.CL}. The existence proof for solitons is contained 
in \S\ref{sec:ES:ExistProof} and relies on a natural condition for the 
super-quadratic growth of the functional $\calP$. In \S\ref{sec:ES:SQCrit} and
\S\ref{sec:ES:GSQCrit} we then discuss the corresponding properties
for the interaction potential $\Phi$. Finally, inspired by the notion of
$\Gamma$-convergence we show in \S\ref{sec:ES.GC} that wave trains
converge to solitons when the periodicity length tends to infinity.
%
%
%
%
%
%
%--------------------------------------------------------------------
\section{%
Variational approach
% via constrained optimisation and improvement dynamics
}%
%\label{sec:overview}
%--------------------------------------------------------------------
%
%
In this section we transform the travelling wave equation into a
fixed point equation for a normalised profile function $W$, and
describe our variational approach to existence results for both wave
trains and solitons. To point our the key ideas we start with more
formal consideration in \S\ref{sec:overview.reform} and
\S\ref{sec:overview.variational},
and postpone the analytical details to \S\ref{sec:tools}. %
%
%
%--------------------------------------------------------------------
\subsection{%
Travelling waves as eigenfunctions of nonlinear integral equations
}%
\label{sec:overview.reform}
%--------------------------------------------------------------------
%
In what follows we assume that the periodicity length of wave trains
is given by $2L$ with $0<L<\infty$, and regard the corresponding
profile functions ${R}$ and ${V}$ as defined on
$\ccinterval{-\Len}{\Len}$. Moreover, we identify the soliton case
with $\Len=\infty$ by considering ${R}$ and ${V}$ as functions on
$\ccinterval{-\infty}{\infty}$, or, equivalently, as functions on
the Alexandrov compactification of $\Rset$. In other words, in both cases we impose
the 'boundary conditions'
$R\at{L}=R\at{-L}$ and  $V\at{L}=V\at{-L}$.
\par%
In what follows we denote by
$\fspaceL^2\at{\ccinterval{-\Len}{\Len}}$ with $L\in\ocinterval{0}{\infty}$ 
the Lebesgue space of all square-integrable functions on $\ccinterval{-\Len}{\Len}$, and if
there is no risk of confusion we write $\fspaceL^2$ instead of
$\fspaceL^2\at{\ccinterval{-\Len}{\Len}}$.
\bigpar
Our first aim is to transform the travelling wave equations for wave
trains and solitons into eigenvalue equations for certain nonlinear
integral operators defined on $\fspaceL^2$. For this purpose we
define two linear averaging operators, and normalise the potential
$\Phi$. More precisely, for given reference distance $r_0$ we define
the potential ${\Phi}_{r_0}$ by
\begin{align*}
{\Phi}_{r_0}\at{r}=\Phi\at{r_0+r}-\Phi\at{r_0}-\Phi^\prime\at{r_0}r
\end{align*}
which is \emph{normalised} via
${\Phi}_{r_0}\at{0}=\Phi_{r_0}^\prime\at{0}=0$ and
$\Phi_{r_0}^{\prime\prime}\at{0}=\Phi^{\prime\prime}\at{r_0}$, and
mention that this normalisation respects the convexity of
$\Phi$. The eigenvalue equations for solitons and wave trains involve the
averaging operators
\begin{align}
\notag%\label{Intro:Def.AvOperator.1}%
\at{\widebar{\calA}_k{W}}\at{\phase}=
\int\limits_{\phase-k/2}^{\phase+k/2}
{W}\at{\tilde\phase}\dint{\tilde\phase},
\end{align}
and
\begin{align}
\notag%\label{Intro:Def.AvOperator.2}%
\widehat{A}_k{W}=
\widebar{\calA}_k{W}-\frac{1}{2\Len}
\int\limits_{-\Len}^{\Len}\widebar{\calA}_k{W}\at{\tilde\phase}\dint\tilde\phase
=%
\widebar{\calA}_k{W}-\frac{k}{2\Len}
\int\limits_{-\Len}^{\Len}{W}\at{\tilde\phase}\dint\tilde\phase.
\end{align}
Notice that the operator $\widebar{\calA}_k$ is well defined and
symmetric on $\fspaceL^2$ for both finite and
infinite $L$, compare Lemma \ref{AP:Lem.AProps} below, whereas
$\widehat{\calA}_k$ is well defined for $\Len<\infty$ only.
%
%
%--------------------------------------------------------------------
\paragraph*{Wave trains and normalisation via mean values}
%--------------------------------------------------------------------
%
%
In order to reformulate the travelling wave equation
\eqref{Intro:TWEqn} for $\Len<\infty$ we introduce the \emph{mean
values} of wave trains by
\begin{align*}
r_{\av}\deq\frac{1}{2\Len}
\int\limits_{-\Len}^{\Len}{R}\at\phase\dint\phase
,\qquad%
v_{\av}\deq\frac{1}{2\Len}
\int\limits_{-\Len}^{\Len}{V}\at\phase\dint\phase.
\end{align*}

\begin{observation}
\label{OV:ScalWT.Obs}%
With the identification
\begin{align}
\label{OV:ScalWT.Trafo}%
{R}\at{\phase-k/2}=r_\av+\nat{\widehat{\calA}_k{W}}\at{\phase}
,\qquad%
{V}\at{\phase}=v_\av+\om{W}\at{\phase},
\end{align}
for some profile function ${W}$ with
$\tfrac{1}{2L}\int_{-L}^L{W}\at\phase\dint\phase=0$  the
integral equation
\begin{align}
\label{OV:ScalWT.Eqn}%
\om^2{W}=\widehat{\calA}_k\Phi_{r_\av}^\prime\nat{\widehat{\calA}_k{W}}.
\end{align}
is equivalent to the wave train equation \eqref{Intro:TWEqn}.
\end{observation}
\begin{proof}
First suppose that ${R}$ and ${V}$ solve $\eqref{Intro:TWEqn}$, and
let ${W}=\om^{-1}\at{{V}-v_\av}$. Then, the first equation in
$\eqref{Intro:TWEqn}$ implies
\begin{align*}
\frac{\dint}{\dint\phase}{R}\at{\phase-k/2}
=
{W}\at{\phase+k/2}-{W}\at{\phase-k/2}=
\frac{\dint}{\dint\phase}\at{\widehat{\calA}_k{W}}\at{\phase},
\end{align*}
and hence
\begin{math}
{R}\at{\phase-k/2}=\nat{\widehat{\calA}_k{W}}\at{\phase}+c_1
\end{math}
for some constant $c_1$. Integrating this with respect to $\phase$
we find $r_\av=0+c_1$, and hence \eqref{OV:ScalWT.Trafo}$_1$. Moreover, 
the second equation in $\eqref{Intro:TWEqn}$ provides
\begin{align*}
\om^2\frac{\dint}{\dint\phase}{W}\at{\phase}
&=%
\Phi^\prime\Bat{r_\av+\nat{\widehat{\calA}_k{W}}\at{\phase+k/2}}-
\Phi^\prime\Bat{r_\av+\nat{\widehat{\calA}_k{W}}\at{\phase-k/2}}
\\&=
\Phi^\prime_{r_\av}\Bat{\nat{\widehat{\calA}_k{W}}\at{\phase+k/2}}-
\Phi^\prime_{r_\av}\Bat{\nat{\widehat{\calA}_k{W}}\at{\phase-k/2}},
\end{align*}
and integration with respect to $\phase$ gives
$\om^2{W}=\widehat{\calA}_k\Phi_{r_\av}^\prime\nat{\widehat{\calA}_k{W}}+c_2$ 
for some constant $c_2$. The condition $\tfrac{1}{2L}\int_{-L}^L{W}\at\phase\dint\phase=0$ 
now implies $c_2=0$ 
and hence \eqref{OV:ScalWT.Trafo}$_2$. Now let a
solution ${W}$ of \eqref{OV:ScalWT.Eqn} be given, and define both
${R}$ and ${V}$ by \eqref{OV:ScalWT.Trafo}. Then, the first equation
from \eqref{Intro:TWEqn} holds by construction, and the second one
follows from  \eqref{OV:ScalWT.Eqn} by differentiation with respect
to $\phase$.
\end{proof}
%
%
%--------------------------------------------------------------------
\paragraph*{Solitons and normalisation via background states}
%--------------------------------------------------------------------
%
For solitons it is much more convenient to base the normalisation on
the \emph{background states}
\begin{align*}
r_{\bgr}\deq\lim_{\phase\to\pm\infty}{R}\at{\phase}
,\qquad%
v_{\bgr}\deq\lim_{\phase\to\pm\infty}{V}\at{\phase},
\end{align*}
because then the normalised profiles are localised.
\begin{observation}
\label{OV:ScalSol.Obs}%
With the identification
\begin{align}
\label{OV:ScalSol.Trafo}%
{R}\at{\phase-k/2}=r_\bgr+\nat{\widebar{\calA}_k{W}}\at{\phase}
,\qquad%
{V}\at{\phase}=v_\bgr+\om{W}\at{\phase},
\end{align}
for some profile function ${W}$ with
$\lim_{\phase\to\pm\infty}{W}\at\phase=0$ the integral
equation
\begin{align}
\label{OV:ScalSol.Eqn}%
\om^2{W}=\widebar{\calA}_k\Phi_{r_\bgr}^\prime\nat{\widebar{\calA}_k{W}}.
\end{align}
is equivalent to the soliton equation \eqref{Intro:TWEqn}.
\end{observation}
\begin{proof}
Similar to that of Observation \ref{OV:ScalWT.Obs}.%
\end{proof}
We can consider the integral equation \eqref{OV:ScalSol.Eqn} also
for $\Len<\infty$, and readily verify that each solution defines a wave train via
\eqref{OV:ScalSol.Trafo}. This normalisation for wave trains seems
to be artificial because for finite $L$ we have no immediate
interpretation of the parameter $r_\bgr$. However, in
\S\ref{sec:ES.GC} we rely on this setting and show that the
$\Len$-periodic solutions ${W}_{L}$ to \eqref{OV:ScalSol.Eqn}
converge to solitons as $L\to\infty$, see Corollary
\ref{GC:Corr.Conv4}, an this implies ${R}_L\at{\pm{L}}\to{r_\bgr}$.
%
%
%--------------------------------------------------------------------
\paragraph*{Parameter dependence of travelling waves}
%--------------------------------------------------------------------
%
The travelling wave equation \eqref{Intro:TWEqn} obeys a simple
scaling symmetry. In fact, the solution set to \eqref{Intro:TWEqn}
is invariant under
\begin{align}
\label{Intro.Scaling}
{R}\at\phase{\,\rightsquigarrow\,}{R}\at{\la\phase},
\quad%
{V}\at\phase{\,\rightsquigarrow\,}{V}\at{\la\phase},
\quad{}%
k{\,\rightsquigarrow\,}\la^{-1}{k},\quad{}\om{\,\rightsquigarrow\,}\la^{-1}\om,
\quad%
\Len{\,\rightsquigarrow\,}\la^{-1}{\Len},
\end{align}
with arbitrary $\la>0$, corresponding to
\begin{align*}
{W}\at\phase{\,\rightsquigarrow\,}\la{W}\at{\la\phase}
\quad\text{with}\quad
r_\bgr{\,\rightsquigarrow\,}r_\bgr,
\quad{v_\bgr}{\,\rightsquigarrow\,}{v_\bgr},\quad
r_\av{\,\rightsquigarrow\,}r_\av,
\quad{v_\av}{\,\rightsquigarrow\,}{v_\av}.
\end{align*}
Up to this scaling there remain four independent parameters for wave
trains. A natural choice for nonlinear potentials is to fix the
length parameter $\Len$, and to regard $r_\av$, $v_\av$, $k$, and
$\om$ as independent parameters, see \cite{DHM06}. In this paper,
however, we prefer to fix the wave number $k$, so that wave trains
are parametrised by $r_\av$, $v_\av$, $\Len$, and the \emph{phase
speed} $\si\deq\om/k$. Accordingly, we parametrize solitons by
$r_\bgr$, $v_\bgr$ and $\si$.
\par%
Notice that the velocity parameters $v_\av$ and $v_\bgr$ do not
appear in the travelling wave equations due to the Galilean
invariance of FPU chains. Both $v_\av$ and $v_\bgr$  are thus
trivial parameters, which however become important when studying
\emph{modulated travelling waves} for which the parameters vary on a
macroscopic scale, compare \cite{DHM06,DH08}.
%
%
%--------------------------------------------------------------------
\subsection{Variational structure for wave trains and solitons}
\label{sec:overview.variational}
%--------------------------------------------------------------------
%
%
%
In view of the scaling and reformulation results from
\S\ref{sec:overview.reform} we start to simplify our setting. In
what follows we consider an arbitrary convex and smooth potential
$\Phi$ normalised by $\Phi\at{0}=\Phi^\prime\at{0}=0$. Moreover, we
restrict to $k=1$, so that the frequency $\om$ equals the phase
speed $\si$, and consider solely the following two averaging
operators
\begin{align}
\label{OV:AvOp}
\at{\widebar{\calA}{W}}\at{\phase}=
\int\limits_{\phase-1/2}^{\phase+1/2}{W}\at{\tilde\phase}
\dint{\tilde\phase}
,\qquad \nat{\widehat{A}{W}}\at{\phase}=
\at{\widebar{\calA}{W}}\at\phase-\frac{1}{2\Len}
\int\limits_{-\Len}^{\Len}{W}\at{\phase}\dint\phase.
\end{align}
\bigpar%
Moreover, in order to point out the similarities between wave trains
and solitons we refer to an abstract averaging operator $\calA$,
which equals either $\widebar{\calA}$ or $\widehat{\calA}$, and
consider the general travelling wave equation
\begin{align}
\label{OV:TWEqn}
\si^2{W}=\calA\Phi^\prime\at{\calA{W}},
\end{align}
with ${W}\in\fspaceL^2\at{\ccinterval{-\Len}{\Len}}$, where $\Len$
might be finite or infinite. In what follows we call $W$ a
\emph{travelling wave } if and only if there exist $\si^2>0$ such
that \eqref{OV:TWEqn} is satisfied. %%
\bigpar
Starting point for the variational formulation is the observation
that \eqref{OV:TWEqn} is the Euler-Lagrange equation to the action
functional
\begin{align*}
\calL\pair{\si^2}{{W}}=\calK\pair{\si^2}{{W}}-\calP\at{{W}},
\end{align*}
where
\begin{align}
\label{OV:Energies} %
\calK\pair{\si^2}{{W}}=
\tfrac{1}{2}\,\si^2\int\limits_{-\Len}^\Len{W}\at{\phase}^2\dint\phase
\qquad\text{and}\qquad
\calP=\int\limits_{-\Len}^\Len\Phi\bat{\at{\calA{W}}\at\phase}\dint\phase
\end{align}
are the \emph{kinetic} and \emph{potential} energy, respectively.
\bigpar
The first rigorous existence result for wave trains and solitons
under super-quadratic growth assumptions for $\Phi$ was given by
Friesecke and Wattis \cite{FW94}. The key idea is (in our notations)
to minimise the $\fspaceL^2$-norm of $W$ under the constraint of
prescribed potential energy, where $1/\si^2$ plays the role of a
Lagrangian multiplier. The existence of corresponding minimisers was
then established by means of Lions' concentration-compactness
principle \cite{Lio84}.
\par
Smets and Willem \cite{SW97} prove the existence of solitons by
showing that for super-quadratic $\Phi$ the functional $\calL$
satisfies the assumptions of (a modified) Mountain Pass Theorem.
Recently, these results could be improved by Schwetlick and Zimmer.
In \cite{SZ07} they require the super-quadratic growth to hold only
asymptotically, so certain double-well potentials are admissible.
\par
Similarly to \cite{SW97}, Pankov and Pfl\"uger \cite{PP00} apply the
Mountain Pass Theorem to wave trains, and pass to the limit
$\Len\to\infty$ by means of concentration compactness. Moreover,
they present a different existence proof for solitons which is based
on the Nehari manifold of $\calL$.
\bigpar
A new idea for convex potentials $\Phi$ was introduced by Filip and
Venakides \cite{FV99} by proposing to maximise the potential energy
$\calP$ under the convex constraint ${W}\in\calB_\ga$ with
\begin{align*}
{\calB_\ga}=\{{W}\in\fspaceL^2\;:\;
\tfrac{1}{2}\norm{{W}}_2^2\leq\ga\}.
\end{align*}
A first advantage of this approach is that $L<\infty$ implies the
functional $\calP$ to be continuous with respect to the weak
topology in $\fspaceL^2$, so that the existence of wave trains
follows from elementary principles of infinite-dimensional convex analysis. 
Secondly, the improvement operator $\calT_\ga$ appears naturally in this context,
and gives rise to effective approximation schemes for wave trains,
see \cite{Her05,DH08} and the numerical simulations below.
\par
Our
method is also based on the constrained
maximization of the potential energy but yields improved results as
it exploits several invariance properties $\calT_\ga$.
%
%
%--------------------------------------------------------------------
\subsubsection*{%
Constrained maximisation of the potential energy via the improvement
operator
}%
%--------------------------------------------------------------------
%
%
%
In view of \eqref{OV:TWEqn} we formally define the
\emph{improvement} operator $\calT_\ga$ by
\begin{align}
\notag%\label{IS:Def.ItOp} %
\calT_{\ga}\ato{{W}}
\deq\frac{\sqrt{2\ga}}{\norm{\partial\calP\ato{{W}}}_2}\partial\calP\ato{W},
\end{align}
where the operator $\partial\calP$ is the G\^{a}teaux derivative of
$\calP$, that means
$\partial\calP\ato{{W}}=\calA\Phi^\prime\at{\calA{W}}$ for all
$W\in\fspaceL^2$. By construction, each fixed point $W$ of
$\calT_\ga$ is a travelling wave with
$\tfrac{1}{2}\norm{{W}}_2^2=\ga$, and vice versa, where the speed is
given by $\si^2=\norm{\partial\calP\ato{{W}}}_2/\sqrt{2\ga}$.
\bigpar%
In the next Section \S\ref{sec:tools} we exploit the convexity of
$\Phi$ and derive the following building blocks for the existence
proof.
\begin{enumerate}
\item
$\partial\calP$ respects the positive cone
\begin{align*}
\calU&\deq\{{W}\in\fspaceL^2
\;:\;%
{W}\at{-\phase_1}={W}\at{\phase_1}
\;\text{and}\;%
{W}\at{\phase_1}\geq{} {W}\at{\phase_2}%
\;\text{for almost all}\;%
0\leq\phase_1\leq\phase_2\leq\Len\},
\end{align*}
which consists of all functions on $\ccinterval{-\Len}{-\Len}$ that
are even and unimodal. Moreover, for $\calA=\widebar{\calA}$ the
operator $\partial\calP$ also respects
\begin{align*}
\calN&\deq\{{W}\in\fspaceL^2
\;:\;%
{W}\at{\phase}\geq0
\;\text{for almost all}\;%
\phase\in\ccinterval{-\Len}{-\Len}\},
\end{align*}
which is the cone of all non-negative functions.
\item
$\calT_\ga$ is well defined on $\calB_\ga\setminus\calM$ and maps
into $\partial\calB_\ga\setminus\calM$, where
\begin{align*}
\calM\deq\big\{{W}\in\fspaceL^2\;:\;\calP\at{{W}}=0\big\}
\end{align*}
is the set of all global minimisers of $\calP$.

\item%
$\calT_\ga$ increases the potential energy, that means
$\calP\at{\calT_\ga{W}}\geq\calP\at{{W}}$ for all $W\notin\calM$,
where equality holds if and only if $W=\calT_\ga\ato{W}$.
\end{enumerate}
We are now able to describe the key principle that provides the
existence of travelling waves.
\begin{theorem}
\label{IS:Cone2.Theo}
Let $\calS\subset{\fspaceL^2}$  be some positive cone that is
invariant under the action of the operator $\partial\calP$. Then,
the set $\calS_\ga\setminus\calM$  with
$\calS_\ga=\calS\cap\calB_\ga$ is invariant under the action of
$\calT_{\ga}$,and each proper maximiser of $\calP$ in $\calS_\ga$ is
a fixed point of $\calT_{\ga}$, and hence a travelling wave.
\end{theorem}
\begin{proof}
The invariance of $\calS_\ga$ is implied by the assumption on
$\calS$ and the properties of $\calT_\ga$.  Now let $W$ be a proper
maximiser. Then,
$\calP\at{\calT_\ga\ato{W}}=\calP\at{W}>\min\calP|_{\calS_\ga}$
implies both $\calT_\ga\ato{W}={W}$ and $W\neq\calM$, and we
conclude that $W$ is in fact a travelling wave with $\si^2>0$.
\end{proof}
In what follows the cone $\calS$ is given by either $\calU$ or
$\calU\cap\calN$. Since these cones are not open in $\fspaceL^2$ the
fact that each maximiser must satisfy the Euler-Lagrange equation
\eqref{OV:TWEqn} with multiplier $\sigma^2$ is not clear a priori
but provided by the invariance of $\calS_\ga$ under $\calT_\ga$.
\par
Theorem \ref{IS:Cone2.Theo} yields only a sufficient condition for
the existence of travelling waves.  In fact, to show that $\calP$
attains its maximum in $\calS_\ga$ is not trivial at all (at least
in the soliton case), and requires a better understanding of the
energy landscape in $\calS_\ga$. In our analysis we follow the
direct approach and show that maximising sequences for $\calP$ are
compact in some appropriate topology in $\fspaceL^2$. More
precisely, for wave trains we use weak compactness, whereas in the
soliton setting we establish the strong compactness for maximising
sequences.
%
%--------------------------------------------------------------------
\subsection{Some pieces of functional analysis}
\label{sec:tools}
%--------------------------------------------------------------------
%
Here we prove the aforementioned properties of the improvement
operator $\calT_\ga$. To this end we rely on following standing
assumptions on the potential $\Phi$.
\begin{assumption}
\label{AP:PotAss}
For given $\gamma>0$ we assume that the interaction potential $\Phi$
has the following properties on the interval
$\ccinterval{-\sqrt{2\ga}}{\sqrt{2\ga}}$.
\begin{enumerate}
\item
(Smoothness) $\Phi$ is at least $\fspaceC^2$,
\item
(Convexity) $\Phi^{\prime\prime}\geq0$,
\item
(Normalisation) $0=\Phi\at{0}=\Phi^\prime\at{0}$ and
$\Phi^{\prime\prime}\at{0}=\beta\geq0$,
\item
(Non-triviality) $\Phi$ does not vanish identically.
\end{enumerate}
\end{assumption}
The restriction to the interval
$\ccinterval{-\sqrt{2\ga}}{\sqrt{2\ga}}$ is natural in our context,
because ${W}\in\calB_\ga$ implies
$\norm{\calA{W}}_\infty\leq\sqrt{2\ga}$, see Lemma
\ref{AP:Lem.AProps}. As a consequence of Assumption \ref{AP:PotAss}
we find
\begin{align}
\label{AP:PotAss.Eqn1}
0\leq\Phi\at{r}\leq\tfrac{1}{2}r^2\at{\beta+\Do{\abs{r}}}
,\qquad
\Phi^\prime\at{-\abs{r}}\leq{0}\leq\Phi^\prime\at{\abs{r}}
\end{align}
for all $r$ with $\abs{r}\leq\sqrt{2\ga}$. Moreover, the
non-triviality condition implies $\calM=\{W:\calP\at{W}=0\}$ and
$\calB_\ga\setminus\calM\neq\{0\}$, and hence each maximiser of
$\calP$ in $\calB_\ga$ is proper.
\bigpar
Within this section, the parameter $L$ can take arbitrary values in
$\ocinterval{0}{\infty}$, and $\fspaceL^p$ and $\fspace{W}^{1,\,p}$
with $1\leq{p}\leq\infty$ denote the usual Lebesgue and Sobolev
spaces on $\ccinterval{-\Len}{\Len}$, where
\begin{align*}
\skp{W_1}{W_2}=
\int\limits_{-\Len}^{\Len}W_1\at\phase{W}_2\at\phase\dint\phase
\end{align*}
gives the inner product in $\fspaceL^2$.
%
%
%--------------------------------------------------------------------
\subsubsection*{%
Properties of the averaging operators $\widebar{\calA}$ and
$\widehat{\calA}$
}%
%--------------------------------------------------------------------
%
%
We summarize some elementary properties of the averaging operators
that are used in the proofs below.
\begin{lemma}
\label{AP:Lem.AProps}
For any $\Len$, the operator
$\widebar{\calA}$ is well defined on $\fspaceL^2$ and has the
following properties:
\begin{enumerate}
\item
$\widebar{\calA}$ maps into $\fspaceL^2\cap{}\fspaceL^\infty$ with
\begin{math}
\norm{\widebar{\calA}{W}}_\infty\leq\norm{{W}}_2
\end{math}
and
\begin{math}
\norm{\widebar{\calA}{W}}_2\leq\norm{{W}}_2
\end{math}.
\item
$\widebar{\calA}$ maps into $\fspaceW^{1,\,2}$ with
\begin{math}
\at{\widebar{\calA}{{W}}}^\prime\at\phase=
{W}\at{\phase+1/2}-{W}\at{\phase-1/2}.
\end{math}
\item
$\widebar{\calA}$ is self-adjoint on $\fspaceL^2$.
\item
If a sequence $\at{{W}_n}_n$ converges weakly in $\fspaceL^2$ to
some limit ${W}_\infty$ , then $\at{\widebar{\calA}{W}_n}_n$
converges strongly in
$\fspaceL^2\nat{\ccinterval{-\tilde{\Len}}{\tilde{\Len}}}$ for each
finite $\tilde{\Len}<\infty$ with $\tilde{\Len}\leq{\Len}$. In
particular, for $\Len<\infty$ the image of each bounded set in
$\fspaceL^2$ under the operator $\widebar{\calA}$ is pre-compact in
$\fspaceL^2$ with respect to the strong topology.
\item
In the wave train case ($\Len<\infty$) the operator
$\widebar{\calA}$ is compact. Moreover, the $m$th eigenvalue
(m=0,\,1,\,2,\,\tdots)
is given by %
\begin{align*}
\varrho_m=\specA\at{\frac{m\pi}{2\Len}},\qquad
\specA\at{\varrho}\deq\varrho^{-1}\sin\at{\varrho},
\end{align*}
and the corresponding eigenspace is spanned by
$\cos\at{\frac{m\pi}{\Len}\cdot}$ and $\sin\at{\frac{m\pi}{\Len}\cdot}$
\item
In the soliton case ($\Len=\infty$) the operator $\widebar{\calA}$
is \emph{not} compact anymore, because it has continuous spectrum
$\mathrm{spec}_{\fspaceL^2}\widebar{\calA}=\{\specA\at\varrho\;:\;\varrho\in\Rset\}$.
In particular, $\mathrm{spec}_{\fspaceL^2}\widebar{\calA}^2=
\ccinterval{0}{1}$.
\end{enumerate}
Moreover, for $\Len<\infty$ we have
$\widehat{\calA}:\fspaceL^2\to{\fspaceL^2}\cap{\fspaceL^\infty}$ with
\begin{math}
\norm{\widehat{\calA}{W}}_\infty
\leq%
\norm{{W}}_2
\end{math}
and
\begin{math}
\norm{\widehat{\calA}{W}}_2
\leq%
\norm{{W}}_2
\end{math}.
\end{lemma}
\begin{proof}
Definition \eqref{OV:AvOp} gives
\begin{align}
\label{AP:Lem.AProps.EqnA}
\widebar{\calA}{V}\at\phase
=%
\int\limits_{-\Len}^{\Len}\widebar{\chi}\at{\phase-s}{V}\at{s}\dint{s}
=%
\widebar{\chi}\at{\phase-\cdot}\ast{V}
\end{align}
where $\widebar{\chi}$ abbreviates the indicator function of the
interval $\ccinterval{-1/2}{1/2}$, and $\ast$ denotes the
convolution operator. H\"older's inequality provides
\begin{math}
\abs{\widebar{\calA}{V}\at\phase}^2
\leq%
\int_{\phase-1/2}^{\phase+1/2}{V}\at{s}^2\dint{s},
\end{math} %
and from this we readily derive the first assertion. The proofs of
the second and the third claim are then straight forward. Now
suppose that ${W}_n\to{W}_\infty$ weakly in
$\fspaceL^2\nat{\ccinterval{-\Len}{\Len}}$, so that
$\widebar{\calA}{W}_n\to\widebar{\calA}{W}_\infty$ point-wise thanks
to \eqref{AP:Lem.AProps.EqnA}, and this implies the strong
$\fspaceL^2$-convergence on each finite interval
$\ccinterval{-\tilde{\Len}}{\tilde{\Len}}$ due to the uniform
$\fspaceL^\infty$-bounds.
\par
Towards the spectral properties of $\widebar{\calA}$ we study how
$\specA$ acts on plane waves. A direct calculation shows that each
plane wave $E_k\at\phase=\phase\mapsto\mhexp{\iu{k}\phase}$
satisfies the eigenvalue equation
\begin{align*}
\widebar{\calA}E_k=\specA\at{k/2}E_k
\end{align*}
pointwise, and this implies the fourth and the fifth assertion.
Finally, for $\Len<\infty$ we have
\begin{align*}
\widebar{\calA}{V}\at\phase=
\widehat{\chi}\at{\phase-\cdot}\ast{V},\qquad
\widehat{\chi}\at{\phase}=\widebar{\chi}\at{\phase}-\tfrac{1}{2\Len}.
\end{align*}
This implies
\begin{math}%
\nabs{\widehat{\calA}{W}\at\phase}^2\leq\norm{\widehat{\chi}}_{\infty}
\norm{\widebar{\chi}\at{\phase-\cdot}{W}}^2_{2}
=%
\nabs{\widebar{\calA}{W}\at\phase}^2
\end{math} %
and in turn the desired properties of $\widehat{\calA}$.
\end{proof}
As a consequence of Definition \eqref{OV:AvOp} and Lemma
\ref{AP:Lem.AProps} we easily find
\begin{align}
\label{AP:Eqn.AKern.1}
\ker_{\fspaceL^2}{\widebar{\calA}}
=%
\Big\{W\in\fspaceL^2\;:\;W\at{\cdot}
=W\at{\cdot+1},\quad\int\limits_{-1/2}^{1/2}W\at\phase\dint\phase=0
\Big\}.
\end{align}
In particular, the kernel of $\widebar{\calA}$ is trivial if either
$\Len$ is irrational, or $\Len=\infty$. Moreover, for $\Len<\infty$
we have
\begin{align}
\label{AP:Eqn.AKern.2}
\ker_{\fspaceL^2}{\widehat{\calA}}
=%
\ker_{\fspaceL^2}\widebar{\calA}\oplus\mathrm{span}\{1\}.
\end{align}
%
%
%--------------------------------------------------------------------
\subsubsection*{Properties of the potential energy functional $\calP$}
%--------------------------------------------------------------------
%
%
%
We rely on Assumption \ref{AP:PotAss} and use standard methods from
convex analysis to prove some properties of $\calP$ and its
derivative. All results are formulated in terms of the abstract
averaging operator $\calA$ and hold both in the wave train and the
soliton case.
\begin{lemma}
\label{AP:Lem.WProps}$\;$ The functional $\calP$ is well-defined,
bounded, continuous and G\^{a}teaux-differentiable on $\calB_\ga$,
and its derivative $\partial\calP=\calA\circ\partial\Phi\circ\calA$
is a monotone operator and maps $\calB_\ga$ continuously into
$\fspaceL^2$. Moreover, for arbitrary ${W}_1,{W}_2\in{\calB}_\ga$ we
have
\begin{align}
\label{AP:Rem.WProps.Eqn2} %
\calP\at{{W}_2}-\calP\at{{W}_1}
\geq%
\tfrac{m}{2}\norm{\calA{W}_2-\calA{W}_1}_2^2+
\skp{\partial\calP\ato{{W}_1}}{{W}_2-{W}_1}.
\end{align}
and
\begin{align}
\label{AP:Rem.WProps.Eqn1} %
\skp{\partial\calP\ato{{W}_2}-
\partial\calP\ato{{W}_1}}{{W}_2-{W}_1}
\geq{m}%
\norm{\calA{W}_2\at\phase-\calA{W}_1\at\phase}_2^2
\end{align}
where the monotonicity constant $m$ is given by
$m=\inf_{\abs{r}\leq\sqrt{2\ga}}\Phi^{\prime\prime}\at{r}\geq\beta$.
\end{lemma}
\begin{proof}
For all ${W}$ in $\calB_\ga$ we have
$\norm{\calA{W}}_\infty\leq\sqrt{2\ga}$, and Assumption
\ref{AP:PotAss} implies
\begin{align*}
\abs{\Phi^\prime\at{r}}\leq{}{C}\abs{r},\qquad
\Phi\at{r}\leq{}\tfrac{1}{2}{C}r^2
\end{align*}
for all $r$ with $\abs{r}\leq\sqrt{2\ga}$, where
\begin{math}
{C}={\sup_{\abs{r}
\leq%
\sqrt{2\ga}}{\abs{\Phi^{\prime\prime}\at{r}}}}
\geq%
\beta
\end{math}. %
Consequently, we find
\begin{align*}
0
\leq%
\calP\at{{W}}
\leq%
\tfrac{1}{2}C\norm{\calA{W}}_2^2
\leq%
\ga\,C,
\qquad%
\norm{\calA{\partial\Phi}\ato{\calA{W}}}_p
\leq{}%
C^p\norm{{W}}_p
\leq{}%
C^p\norm{{W}}_p,
\end{align*}
and all assertions concerning the continuity and boundedness of both
$\calP$ and $\partial\calP$ follow immediately. Now let
${W}_1,\,{W}_2\in\calB_\ga$ be fixed, and notice that the convexity
inequality
\begin{math} %
\at{\Phi^{\prime}\at{r_2}-\Phi^{\prime}\at{r_1}}\at{r_2-r_1}
\geq{m}\at{r_2-r_1}^2
\end{math} %
with $r_i={W}_i\at\phase$ implies \eqref{AP:Rem.WProps.Eqn1} by
integration w.r.t. $\phase$. To prove \eqref{AP:Rem.WProps.Eqn2},
let $\eta\in\ccinterval{0}{1}$ and consider
${W}\at{\eta}\deq\at{1-\eta}{W}_1+\eta{W}_2\in\calB_\ga$ as well as
\begin{align*}
p\at{\eta}\deq\calP\at{{W}\at{\eta}}.
\end{align*}
The function $p$ is well-defined and differentiable with respect to
$\eta$, and using \eqref{AP:Rem.WProps.Eqn1} we find
\begin{align*}
\frac{\dint{}}{\dint{\eta}}p\at{\eta}
&=%
\skp{\partial\calP\ato{{W}\at{\eta}}}{{W}_2-{W}_1}
\\&\geq%
\eta^{-1}\skp{\partial\calP\ato{{W}\at{\eta}}-
\partial\calP\ato{{W}_1}}{\eta{W}_2-\eta{W}_1}
+\skp{\partial\calP\ato{{W}_1}}{{W}_2-{W}_1}
\\&=%
\eta^{-1}\skp{\partial\calP\ato{{W}\at{\eta}}-
\partial\calP\ato{{W}_1}}{{W}\at{\eta}-{W}_1}
+\skp{\partial\calP\ato{{W}_1}}{{W}_2-{W}_1}
\\&\geq%
\eta^{-1}{m}\norm{\calA{W}\at{\eta}-\calA{W}_1}_2^2
+\skp{\partial\calP\ato{{W}_1}}{{W}_2-{W}_1}
\\&\geq%
\eta{m}\norm{\calA{W}_2-\calA{W}_1}_2^2
+\skp{\partial\calP\ato{{W}_1}}{{W}_2-{W}_1}.
\end{align*}
Finally, we integrate the last estimate from $\eta=0$ to $\eta=1$, and
this gives \eqref{AP:Rem.WProps.Eqn2}.
\end{proof}
The convexity of $\calP$ implies that each trivial travelling wave
with $\si^2=0$ must belong to $\calM$, the set of all minimisers of
$\calP$.
\begin{remark}
\label{AP:Rem.WProps.3}%
We have $\partial\calP\ato{{W}}\neq0$ for all
${W}\in\calB_\ga\setminus\calM$.
\end{remark}
\begin{proof}%
Assume there exists some ${W}\in\calB_\ga\setminus\calM$ with
$\partial\calP\ato{{W}}=0$. Then, \eqref{AP:Rem.WProps.Eqn1} with
${W}_2=0$ and ${W}_1={W}$ provides $\calA{W}=0$ and hence
$\calP\at{{W}}=0$, which is a contradiction.
\end{proof}%
Notice that for non-degenerate potentials $\Phi$ with $\Phi\at{r}>0$
for all $r\neq0$ we have $\calM=\ker{\calA}$ , where $\ker{\calA}$
is given in \eqref{AP:Eqn.AKern.1} and \eqref{AP:Eqn.AKern.2}.
%
%--------------------------------------------------------------------
\subsubsection*{Properties of the improvement operator $\calT_\ga$}
%--------------------------------------------------------------------
%
First we show that the cones $\calU$ and $\calN$ are invariant under
the action of both $\widebar{\calA}$ and $\Phi^\prime$. Here again
the convexity of $\Phi$ enters as it guarantees that $\Phi^\prime$
increases monotonically.
\begin{lemma}
%\label{AP:InvCones}
%
The cones $\calU$ and $\calN$ are convex, closed under weak and
strong convergence in $\fspaceL^2$, and are invariant under the
action of $\partial\calP$ and $\widebar{\calA}$. Moreover, $\calU$
is also invariant under the action of $\widehat{\calA}$ (for
$\Len<\infty$).
\end{lemma}
\begin{proof}
The only non-trivial assertion is the invariance of $\calU$ under
$\widebar{\calA}$. To prove this, we fix ${W}\in\calU$ and consider
${Y}=\tfrac{\dint}{\dint\phase}\widebar{\calA}{W}$ with
${Y}\at\phase={W}\at{\phase+1/2}-{W}\at{\phase-1/2}$ thanks to Lemma
\ref{AP:Lem.AProps}. This function is odd as
${W}\at\phase={W}\at{-\phase}$ implies
\begin{align*}
{Y}\at{-\phase}={W}\at{-\phase+1/2}-{W}\at{-\phase-1/2}=
{W}\at{\phase-1/2}-{W}\at{\phase+1/2}=-{Y}\at{\phase}.
\end{align*}
Hence it remains to show that ${Y}\at\phase\le0$ for all $\phase\ge0$,
which is equivalent to
\begin{align}
\notag%\label{AP:InvCones.Est1}
{W}\at{\phase+1/2}\leq{W}\at{\phase-1/2},\quad\phase\geq0.
\end{align}
For $1/2\leq\phase\leq{\Len}-1/2$ this estimate follows from
$0\leq\phase-1/2\leq\phase+1/2$. Moreover, for $0\leq\phase\leq1/2$
it holds thanks to ${W}\at{-\phase+1/2}={W}\at{\phase-1/2}$ and
$0\leq-\phase+1/2\leq\phase+1/2$, and for
$\Len-1/2\leq\phase\leq\Len$ it is a consequence of
${W}\at{\phase+1/2}={W}\at{2\Len-\phase-1/2}$ and
$2\Len-\phase-1/2\geq\phase-1/2$.
\end{proof}
\begin{lemma}
\label{IS:Lemma.TProps} %
The operator $\calT_{\ga}$ maps $\calB_\ga\setminus\calM$
continuously into $\partial\calB_\ga\setminus\calM$ and satisfies
\begin{align}
\label{IS:Lemma.TProps.Eqn1} %
\calP\at{\calT_\ga\ato{{W}}}-\calP\at{{W}}\geq\tfrac{1}{2}m
\norm{\calA\calT_\ga\ato{{W}}-\calA{W}}_2^2
\end{align}
for all ${W}\in\calB_\ga\setminus\calM$. Moreover, the equality
sign holds if and only if ${W}$ is a fixed point of $\calT_\ga$.
\end{lemma}
\begin{proof}
Lemma \ref{AP:Lem.WProps} and Remark \ref{AP:Rem.WProps.3} imply
that $\calT_{\ga}$ is well defined and continuous on
$\calB_\ga\setminus\calM$. Moreover,
$\norm{\calT_{\ga}\ato{{W}_2}}_2=\sqrt{2\ga}$ holds by definition.
Now let ${W}_1\in\calB_\ga\setminus\calM$ be fixed, and set
${W}_2\deq\calT_\ga\ato{{W}_1}$ and
$\sigma^2_2\deq\norm{\partial\calP\ato{{W}_1}}_2/\sqrt{2\ga}>0$.
Hence, $\sigma^2_2{W}_2=\partial\calP\ato{{W}_1}$, and from
\eqref{AP:Rem.WProps.Eqn2} we infer that
\begin{align*}
\calP\at{{W}_2}-\calP\at{{W}_1}%
-\tfrac{1}{2}{m}\norm{\calA{W}_2-\calA{W}_1}_2^2
&\geq\sigma_2^{-2}\skp{{W}_2}{{W}_2-{W}_1}
\\&\geq%
\sigma_2^{-2}\at{\norm{{W}_2}_{2}^2-\norm{{W}_2}_{2}\norm{{W}_1}_{2}},
\end{align*}
which gives \eqref{IS:Lemma.TProps.Eqn1} due to
$\norm{{W}_1}_2\leq\norm{{W}_2}_2=\sqrt{2\ga}$. Moreover,
we find an equality sign in the second estimate if and only if
$\norm{{W}_2}^2_2=\norm{{W}_1}^2_2=\skp{{W}_1}{{W}_2}$,
that means if and only if ${W}_1={W}_2$.
\end{proof}
With Lemma \ref{IS:Lemma.TProps} we have derived all ingredients
that we had used in the proof of Theorem \ref{IS:Cone2.Theo}.
%
%--------------------------------------------------------------------
\section{Wave Trains}
%\label{sec:WT}
%--------------------------------------------------------------------
%
As a first application of Theorem \ref{IS:Cone2.Theo} we establish
the existence of wave trains in \S\ref{sec:WT.existence} and proceed
with some comments on the numerical computation of wave trains.
Afterwards we study the complete localisation of wave trains in
\S\ref{sec:WT.CL}.
%
%--------------------------------------------------------------------
\subsection{Existence results}
\label{sec:WT.existence}%
%--------------------------------------------------------------------
%
Our first existence result concerns wave trains that are
renormalised via their mean values. This corresponds to
$\Len<\infty$, $\calA=\widehat{\calA}$, and $\calS=\calU$.
\begin{theorem}
\label{WT:ExistTheo.A}%
For each $L<\infty$ and $\ga>0$ there exist a unimodal and even wave
train $W$ such that $\tfrac{1}{2}\norm{W}_2^2=\ga$ and
\begin{math}
\si^2{W}=\widehat{\calA}\Phi^\prime\nat{\widehat{\calA}{W}}
\end{math}
for some $\si^2>0$.
\end{theorem}
\begin{proof}
With respect to the weak topology in $\fspaceL^2$, the functional
$\calP$ is continuous and the set $\calS_\ga=\calU\cap\calB_\ga$ is
compact. Hence there exist a maximiser $W$, which is moreover
proper, that means
$\sup\calP|_{\calS_\ga}=\calP\at{W}>0=\min\calP|_{\calS_\ga}$,
because $\Phi$ is non-trivial. The desired result now follows from
Theorem \ref{IS:Cone2.Theo}.
\end{proof}
In view of Observation \ref{OV:ScalWT.Obs} and the scaling
\eqref{Intro.Scaling} we infer that Theorem \ref{WT:ExistTheo.A}
implies the existence of a four-parameter family of solutions
$\pair{R}{V}$ to the original travelling wave equation
\eqref{Intro:TWEqn} with fixed $L$. This family is parametrised by
$r_\av$, $v_\av$, $k$, and $\ga$, and for nonlinear potentials we
can moreover expect (at least locally) that $\ga$ can be replaced by
$\om$.
\par
Similar existence result for wave trains in convex FPU chains are
proven in \cite{FV99,DHM06}, but provide only
$\calW\in\partial\calB_\ga$. Our method improves these results as it
establishes the existence of wave trains with the additional
property $W\in\calU$, which in turn implies $R\in\calU$ and
$\si{V}\in\calU$. This sheds light on some observations from
\cite{DH08,HR08a}: The traces of travelling waves found in the
numerical simulations of initial value problems for
\eqref{Intro.FPU1} typically encircle convex sets in the
$\pair{r}{v}$-plane, compare the right picture in Figure 
\ref{Fig:DispersiveShock}. In
particular, these curves have exactly two  extrema in both the
$r$-direction and the $v$-direction, and hence they correspond to
unimodal profile functions $R$ and $V$.
\bigpar
We proceed with some remarks concerning the uniqueness of wave
trains. The norm constraint $\tfrac{1}{2}\norm{W}_2^2=\ga$ alone is
not sufficient for uniqueness as the travelling wave equation is
invariant under shifts in $\phase$. Moreover, since the set of all
$2\tilde{L}$-periodic functions, with $m\tilde{\Len}=\Len$ for some
$m\in\Nset$, is invariant under the action of $\calT_\ga$, we can
construct a whole family of wave trains satisfying the norm
constraint.
\par
From these considerations we conclude that any uniqueness result for
wave trains must prescribe further properties of the profile
function $W$. Motivated by numerical simulations we conjecture, that
for each $\ga$ there exists exactly one travelling wave with
$W\in\calU$ and $\tfrac{1}{2}\norm{W}_2^2=\ga$, but we are not able
to prove this conjecture.
\bigpar
Finally, we use the same arguments as in the proof of Theorem
\ref{WT:ExistTheo.A} to derive a similar existence result for wave
trains in the setting $\Len<\infty$, $\calA=\widebar{\calA}$, and
$\calS=\calU\cap\calN$.
\begin{lemma}
\label{WT:ExistTheo.B}%
For each $L<\infty$ and $\ga>0$ there exist a wave train
$W\in\cal{U}\cap\cal{N}\cap\partial\calB_\ga$ such that
\begin{math}
\si^2{W}=\widebar{\calA}\Phi^\prime\nat{\widebar{\calA}{W}}
\end{math}
for some $\si^2>0$.
\end{lemma}
%
%
%
%
%--------------------------------------------------------------------
\paragraph*{Numerical computation of wave trains}
%--------------------------------------------------------------------
%
It is natural to use the improvement dynamics
\begin{align}
\label{Num.Dynamics}
W\in\calS_\ga\mapsto\calT_\ga\ato{W}\in\calS_\ga
\end{align}
for the approximation of wave trains, and a corresponding discrete
scheme is readily derived and implemented. It was proven in
\cite{Her05} that the orbits generated by \eqref{Num.Dynamics} are
compact in the strong $\fspaceL^2$ topology, but from a theoretical
point of view this result remains unsatisfactory due to the lack of
uniqueness. So it is neither clear that maximisers of $\calP$ in
$\calS_\ga$ are unique, nor that all fixed points of $\calT_\ga$ are
(global) maximisers.
\par%
\begin{figure}[ht!]%
  \centering{%
  \includegraphics[width=0.3\textwidth, draft=\figdraft]%
  {\figfile{nf_profile_1}}%
  \hspace{0.025\textwidth}%
  \includegraphics[width=0.3\textwidth, draft=\figdraft]%
  {\figfile{nf_profile_3}}%
  \hspace{0.025\textwidth}%
  \includegraphics[width=0.3\textwidth, draft=\figdraft]%
  {\figfile{nf_profile_5}}%
  \\%
  \includegraphics[width=0.3\textwidth, draft=\figdraft]%
  {\figfile{nf_profile_7}}%
  \hspace{0.025\textwidth}%
  \includegraphics[width=0.3\textwidth, draft=\figdraft]%
  {\figfile{nf_profile_9}}%
  \hspace{0.025\textwidth}%
  \includegraphics[width=0.3\textwidth, draft=\figdraft]%
  {\figfile{nf_profile_11}}%
  }%
  \caption{%
    Profile functions ${W}$ for several values of
    $\ga$ with $\Len=2$ and $\Phi$ as in \eqref{Num:NestFam.Pot}.
      }%
  \label{Fig:nf_profiles}%
  \bigskip%
  \centering{%
  \includegraphics[width=0.45\textwidth, draft=\figdraft]%
  {\figfile{nf_all_traces}}%
  \hspace{0.025\textwidth}%
  \includegraphics[width=0.45\textwidth, draft=\figdraft]%
  {\figfile{nf_zoom_traces}}%
  }%
  \caption{%
    Traces for the wave trains from Figure \ref{Fig:nf_profiles}.
  }%
  \label{Fig:nf_all_traces}%
\end{figure}%
In numerical simulations, however, we found \eqref{Num.Dynamics} to
have good properties. For a wide class of potentials we observed
rapid convergence to a unique limit independent of the chosen
initial data. In Figure \ref{Fig:nf_profiles} we present the
numerically computed profiles $W$ for different values of $\ga$ with
$\calA=\widebar{\calA}$ and %
\begin{align}
\label{Num:NestFam.Pot}
\Phi\at{r}=\cosh{\at{r}}-1.
\end{align}
For small $\ga$ we can approximate the potential by the harmonic one
$\Phi_\harm\at{r}=\Phi^{\prime\prime}\at{0}r^2$, and hence the
profile $W$ is close to a rescaled plane wave. For increasing $\ga$,
however, the nonlinearity dominates and the profile function becomes
tighter. Figure \ref{Fig:nf_all_traces} shows the corresponding
\emph{traces} in the $\pair{r}{v}$-plane, these are the curves
\begin{align*}
\phase\mapsto\bpair{\widebar{\calA}W\at\phase}{\si{W}\at\phase}
\cong\bpair{R\at\phase}{V\at\phase},
\end{align*}
compare Observation \eqref{OV:ScalSol.Obs}. Surprisingly we find a
nested family of curves, that mean the traces for different values
of $\ga$ do not intersect but fill out a convex region. We are not
able to prove this observation but mention that a similar phenomenon
occurs when FPU chains generate dispersive shocks, see
\cite{AMSMSP:DHR,HR08a}.
%
%
%--------------------------------------------------------------------
\subsection{Complete localisation of wave trains}
\label{sec:WT.CL}
%--------------------------------------------------------------------
%
%
It is well known for strongly nonlinear potentials that in certain
limits the wave trains (and even the solitons) localize completely,
in the sense that -- under a suitable rescaling -- the profile
functions ${W}$ converge to the indicator function of an interval
plus a constant background state. Such profile functions are, up to
renormalisation, equal to the profile functions of travelling waves
in the hard-sphere model for the atomic chain, in which all atomic
interactions are described by elastic collisions. Thus the effect of
localisation can often be linked in a natural way to the high energy
limit of travelling waves. For more details we refer to
\cite{Tod81,FM02,Her05,DHM06}.
\bigpar
In this section we discuss the localisation  phenomenon in our
context, and aim to derive a localisation criterion for wave trains.
To keep the presentation simple we solely consider non-negative and
unimodal profile functions, that means we investigate the
localisation of solutions to \eqref{OV:TWEqn} with $L<\infty$,
$\calA=\widebar{\calA}$, and $\calS=\calU\cap\calN$. Moreover, for
our purpose it is sufficient to assume that the localised limit
profile is given by
\begin{align}
\label{HSL:WCL.Def}
{W}_{\CL}\at\phase=%
\chi_{\ccinterval{-\tfrac{1}{2}}{\tfrac{1}{2}}}\at\phase=%
\left\{%
\begin{array}{ll}%
1&\text{if $\abs{\phase}\leq\tfrac{1}{2}$,}
\\%
0&\text{if $\abs{\phase}>\tfrac{1}{2}$,}
\end{array}
\right.%
\end{align}
see Figure \ref{Fig:cl_profiles}. It is easy to check that this
profile satisfies
\begin{align}
\label{HSL:HSProf.Props}
\norm{{W}_{\CL}}_2=1,\quad%
\nat{\widebar{\calA}{W}_{\CL}}\at{\phase}=\max\{1-\abs{\phase},0\},\quad%
\calP\at{{W}_{\CL}}=2\int\limits_0^1\Phi\at{s}\dint{s}.
\end{align}
\bigpar
In what follows we consider sequences $\at{\Phi_n}_n$ of rescaled
potentials, where each $\Phi_n:\ccinterval{0}{1}\to\Rset$ satisfies
Assumption \ref{AP:PotAss}. Moreover, we refer to a sequence of
profile functions ${{W}}_n\subset{\fspaceL^2}$ as a
\emph{corresponding sequence of maximisers}, if ${{W}}_n$ is a
maximiser of $\calP_n$ in $\calS_{1/2}$ for each $n$, where
$\calP_n$ is the potential energy functional \eqref{OV:Energies}
corresponding to $\Phi_n$.
\begin{figure}[ht!]%
  \centering{
  \includegraphics[width=0.7\textwidth, draft=\figdraft]%
  {\figfile{cl_profiles}}%
  }%
  \caption{%
    The functions ${W}_{\CL}$ and $\widebar{\calA}{W}_{\CL}$.
      }%
  \label{Fig:cl_profiles}%
\end{figure}%
\par
We say, a sequence of such potentials $\at{\Phi_n}_n$ has the
\emph{complete localisation property on $\ccinterval{0}{1}$} if any
corresponding sequence of maximisers converges strongly in
$\fspaceL^2$ to ${W}_\CL$. Our main result in this section is a
necessary condition for the complete localisation of wave trains and is
implied by the following observation.
\begin{lemma}
\label{CL:Lemma0}%
We have
\begin{math}
\norm{\widebar{\calA}{W}}_\infty
<%
\norm{\widebar{\calA}{W}_\CL}_\infty=1
\end{math} %
for any ${W}\in\calS_{1/2}$ with
${W}\neq{W}_{\CL}$.
\end{lemma}
\begin{proof}
${W}\in\calU\cap\calN$ implies
$\norm{\widebar{\calA}{W}}_\infty=\at{\widebar{\calA}{W}}\at{0}$
and H\"older's inequality provides
\begin{align}
\at{\widebar{\calA}{W}}^2\at{0}
&=%
\at{\int\limits_{-1/2}^{1/2}{W}\at{\phase}\dint\phase}^2
=%
\at{\int\limits_{-1/2}^{1/2}{W}_{\CL}\at{\phase}
{W}\at{\phase}\dint\phase}^2
\notag%
\\&\leq%
\label{CL:Lemma0.Eqn1}
\at{\int\limits_{-1/2}^{1/2}{W}_{\CL}\at{\phase}^2\dint\phase}
\at{\int\limits_{-1/2}^{1/2}{W}\at{\phase}^2\dint\phase}
=%
\int\limits_{-1/2}^{1/2}{W}\at{\phase}^2\dint\phase
\\&\leq%
\label{CL:Lemma0.Eqn2} {1}=
\at{\widebar{\calA}{W}_{\CL}}^2\at{0}.
\end{align}
Moreover, the estimate in \eqref{CL:Lemma0.Eqn1} is strict unless
there exist a constant $c$ such that
\begin{align}
\label{CL:Lemma0.Eqn3}
{W}\at\phase={c}{W}_{\CL}\at\phase \quad\text{for
almost all $\phase\in\ccinterval{-\tfrac{1}{2}}{\tfrac{1}{2}}$},
\end{align}
whereas the estimate in \eqref{CL:Lemma0.Eqn2} is strict except for
\begin{align}
\label{CL:Lemma0.Eqn4}
\int\limits_{-1/2}^{1/2}{W}\at{\phase}^2\dint\phase=1.
\end{align}
Now suppose that both \eqref{CL:Lemma0.Eqn3} and
\eqref{CL:Lemma0.Eqn4} are satisfied. Then we have $c=1$, and the
norm constraint $\norm{{W}}_2^2\leq{1}$ implies
${W}\at\phase=0={W}_\CL\at\phase$ for almost all
$\phase\notin\ccinterval{-\tfrac{1}{2}}{\tfrac{1}{2}}$, and hence
${W}={W}_{\CL}$.
\end{proof}
\begin{lemma}
\label{CL:Lemma1}%
The sequence $\at{\Phi_n}_n$  has the complete localisation property
on $\ccinterval{0}{1}$ provided that the following two conditions
are satisfied.
\begin{enumerate}
\item $\calP_n\at{{W}_{\CL}}=1$ for all $n$.
\item
$\Phi_n$ converges uniformly and essentially monotonically to $0$ on
each interval $\ccinterval{0}{r_0}$ with $0<r_0<1$. That means for
any $r_0$ we have $\sup_{0\leq{r}\leq{r_0}}\Phi_{n}\at{r}\to0$ as
$n\to\infty$, and there exists $n_0\at{r_0}$ such that
\begin{align*}
0\leq\Phi_{n_2}\at{r}\leq\Phi_{n_1}\at{r}
\end{align*}
for all $n_2>n_1>n_0\at{r_0}$ and all $0\leq{r}\leq{r_0}$.
\end{enumerate}
\end{lemma}
\begin{proof}
First we assume additionally that ${{W}}_n\to{{W}}_\infty$ weakly in
$\fspaceL^2$, and suppose for contradiction that
${{W}}_\infty\neq{W}_\CL$. Then,
$\norm{\widebar{A}{{W}}_\infty}_\infty<1$ thanks to Lemma
\ref{CL:Lemma0}, and since
$\widebar{A}{{W}}_n\to\widebar{A}{{W}}_\infty$ point-wise and
$\norm{\widebar{A}{{W}}_n}_\infty=\at{\widebar{A}{{W}}_n}\at{0}$ we
find some $0<r_0<1$ such that
$\norm{\widebar{A}{W}_n}_\infty\leq{}r_0$ for almost all $n$.
Therefore,
\begin{math}
\calP_{n_1}\at{{{W}}_{n_2}}
\geq%
\calP_{n_2}\at{{{W}}_{n_2}}
\end{math} %
holds for all $n_2\geq{n_1}$ and all sufficiently large $n_1$,
and since each $W_n$ is a maximisers for $\calP_n$ we also have
\begin{math}
\calP_{n_2}\at{{{W}}_{n_2}}
\geq{}%
\calP_{n_2}\at{{{W}}_{\CL}}\geq{1}.
\end{math} %
We conclude that
\begin{math}
\calP_{n_1}\at{{{W}}_{n_2}}
\geq{1}
\end{math} %
and passing to the limits $n_2\to{\infty}$ and
$n_1\to{\infty}$ we
obtain
\begin{math}
\liminf_{n\to\infty}\calP_{n}\at{{{W}}_{\infty}}
\geq%
1.
\end{math} %
However, $\norm{\widebar{\calA}{W}_\infty}_\infty\leq{r_0}<1$
implies $\lim_{n\to\infty}\calP_{n}\at{{{W}}_{\infty}}=0$, the
desired contradiction. The result obtained so far implies that
${W}_\CL$ is the unique accumulation point of a maximising sequence,
and this yields the weak convergence to ${W}_\CL$ for any maximising
sequence. Finally, the strong convergence follows from
$1={\norm{{W}_\CL}_2}=\norm{{W}_n}_2$ for all $n$.
\end{proof}
Notice that the two conditions from Lemma \ref{CL:Lemma1} imply
$\Phi_n\at{r}\to0$ for all $0\leq{r}<1$, but $\Phi_n\at{1}\to\infty$
as $n\to\infty$, and for this reason it is not clear whether or not
the complete localisation (convergence of maximisers) implies the
convergence of maxima. The simulations from Figure
\ref{Fig:hs_mp_fields}, however, provide evidence for
$\liminf\limits_{n\to\infty}\calP_n\at{{W}_n}>1=
\lim\limits_{n\to\infty}\calP_n\at{{W}_\CL}$.
\bigpar%
Our first application concerns the maximiser for homogeneous
potentials of large degree. For an illustration we refer to the
numerical results in Figure \ref{Fig:hs_mp_profiles} and
\ref{Fig:hs_mp_fields}.

\begin{example}
\label{HSL:Example.1}%
The family of potentials $\Phi_q\at{r}=\tfrac{q+1}{2}r^q$ with $c>0$
and $q>2$ has the complete localisation property on
$\ccinterval{0}{1}$ for $q\to\infty$.
\end{example}
\begin{proof}
$\calP_q\at{{W}_\CL}=1$ follows from \eqref{HSL:HSProf.Props} by a
direct computation, and for fixed $0<r_0<1$ we choose $q_0$ such
that $1+\at{q_0+1}\ln{r_0}<0$. Then we find
$\partial_q\at{\tfrac{q+1}{2}r^q}=\frac{1}{2}r^q\at{1+\at{q+1}\ln{r}}<0$
for all $0\leq{r}<r_0$ and $q>q_0$.
\end{proof}
\begin{figure}[!ht]%
  \centering{%
  \includegraphics[width=0.3\textwidth, draft=\figdraft]%
  {\figfile{cl_mp_profile_4}}%
  \hspace{0.025\textwidth}%
  \includegraphics[width=0.3\textwidth, draft=\figdraft]%
  {\figfile{cl_mp_profile_6}}%
  \hspace{0.025\textwidth}%
  \includegraphics[width=0.3\textwidth, draft=\figdraft]%
  {\figfile{cl_mp_profile_10}}%
  \\%
  \includegraphics[width=0.3\textwidth, draft=\figdraft]%
  {\figfile{cl_mp_profile_20}}%
  \hspace{0.025\textwidth}%
  \includegraphics[width=0.3\textwidth, draft=\figdraft]%
  {\figfile{cl_mp_profile_50}}%
  \hspace{0.025\textwidth}%
  \includegraphics[width=0.3\textwidth, draft=\figdraft]%
  {\figfile{cl_mp_profile_100}}%
  }%
  \caption{%
    Profile functions ${W}$ with
    $\ga=\tfrac{1}{2}$, $\Len=2$, and potentials $\Phi_q$
    as in Remark \ref{HSL:Example.1}. This example
    describes the wave trains for homogeneous
    potentials in the limit of increasing degree.
      }%
  \label{Fig:hs_mp_profiles}%
  \bigskip%
  \centering{%
  \includegraphics[width=0.3\textwidth, draft=\figdraft]%
  {\figfile{cl_mp_speed}}%
  \hspace{0.025\textwidth}%
  \includegraphics[width=0.3\textwidth, draft=\figdraft]%
  {\figfile{cl_mp_energy}}%
  }%
  \caption{%
    Speed $\si$ and potential energy $\calP\at{{W}}$ versus
    $q$ for the wave trains from Figure \ref{Fig:hs_mp_profiles}.
      }%
  \label{Fig:hs_mp_fields}%
\end{figure}%
\bigpar
The second candidate for the complete localisation property is
related to the limit $\ga\to\infty$ for fixed potential $\Phi$. For
this purpose we consider the rescaled potentials
\begin{align}
\label{HSL:RenormPot}%
\Phi_\ga\at{r}\deq\frac{\Phi\at{\sqrt{2\ga}r}}
{2\int\limits_0^1\Phi\at{\sqrt{2\ga}s}\dint{s}},
\end{align}
with corresponding energy functionals $\calP_\ga$, and notice that
the two optimisation problems
\begin{align*}
\calP\to\max{}\;\text{on}\;\calS_{\ga}\qquad\qquad\text{and}\qquad\qquad
\calP_\ga\to\max{}\;\text{on}\;\calS_{1/2},
\end{align*}
are equivalent due to
${W}\in\calS_\ga\Leftrightarrow{W}/\sqrt{2\ga}\in\calS_{1/2}$.
\begin{example}
%\label{HSL:Example.2}%
%
For $\Phi$ as in \eqref{Num:NestFam.Pot} the rescaled potentials
$\Phi_\ga$ from \eqref{HSL:RenormPot} have the complete localisation
property on $\ccinterval{0}{1}$ for $\ga\to\infty$.
\end{example}
\begin{proof}
$\calP_\ga\at{{W}_\CL}=1$ holds by construction, and for each
$0<r_0<1$ one can find $\ga_0$ such that
$\partial_\ga\Phi_\ga\at{r}<0$ for all $0\geq{r}\geq{r_0}$ and
$\ga\geq\ga_0$.
\end{proof}
More generally, the family $\at{\Phi_{\ga}}_\ga$ can be expected to
have the complete localisation property for $\ga\to\infty$ provided
that $\Phi$ grows faster than every polynomial. The super-polynomial
growth condition is necessary as for every homogeneous potential of
degree $q$ we have $\Phi_\ga\equiv\Phi_1$. This reflects the the
homogeneous scaling
\begin{align*}
{W}\rightsquigarrow\la{W}
,\quad%
{\sigma}^2\rightsquigarrow\la^{q-2}\sigma^2.
\end{align*}
and shows that the wave trains for homogeneous or polynomial
potentials do not localise in the limit $\ga\to\infty$.
%
%
%--------------------------------------------------------------------
\section{Solitons}%
%\label{sec:ES}%
%--------------------------------------------------------------------
%
This section we study soliton solutions to \eqref{OV:TWEqn}, that
means we set
\begin{align*} %
\Len=\infty,
\qquad%
\calA=\widebar{\calA},
\qquad%
\calS=\calU\cap\calN.
\end{align*}%
Moreover, we assume that the potential energy $\calP$ is genuinely
super-quadratic (see Definition \ref{ES:SQ.Def} below) and show that
for each $\ga>0$ there exists a maximiser of $\calP$ in
$\calS_\ga=\calU\cap\calN\cap\calB_\ga$, which is a soliton
according to Theorem \ref{IS:Cone2.Theo}.
\bigpar%
In what follows we set
\begin{align}
\notag%\label{AP:Def.Sup}%
P\at\gamma\;\deq\sup\limits_{{W}\in\calS_\ga}\calP\at{{W}},
\end{align}
and in order to compare with the harmonic case we introduce
\begin{align}
\label{AP:Def.Harm.Sup}
P_\harm\at\gamma\;\deq\sup\limits_{{W}\in\calS_\ga}\calP_\harm\at{{W}},
\end{align}
where
\begin{align*}
\calP_\harm\at{{W}}
=%
\int\limits_{-\Len}^{\Len}
\Phi_\harm\bat{\at{\calA{W}}\at{\phase}}\dint\phase
=%
\tfrac{1}{2}\beta\norm{\calA{W}}_2^2
\end{align*}
denotes the energy functional corresponding to
$\Phi_\harm\at{r}=\tfrac{1}{2}\beta{}r^2$.
\par%
First we show $P_\harm\at\ga=\beta\ga$ by studying the maximizing
sequence $\at{{U}_n}_{n}\subset\calS_\ga$ defined by
\begin{align}
\label{AP:Def.Harm.MaxSequ}%
{U}_n\at\phase=\left\{\begin{array}{lcl}
\tfrac{\sqrt{2\ga}}{\sqrt{n}}
\cos\at{\frac{\pi}{2n}\phase}
&\text{for}&%
\abs\phase\leq{n},%
\\%
0&\text{for}&\abs\phase\geq{n}.%
\end{array}\right.
\end{align}
\begin{lemma}
\label{AP:Lemma.MaxSequ}

We have
\begin{math}
\beta\gamma\geq\calP_\harm\at{{U}_n}
\geq%
\beta\gamma\at{1-\nDO{n^{-2}}}
\end{math} %
for all $n$, and hence
\begin{align*}
P_\harm=\sup\calP_\harm|_{\calS_\ga}=
\sup\calP_\harm|_{\calB_\ga}=\be\gamma.
\end{align*} %
\end{lemma}
\begin{proof}
A direct calculation shows $\tfrac{1}{2}\norm{{U}_n}_2^2=\ga$, as
well as
\begin{align*}
\at{\calA{U}_n}\at\phase=\left\{\begin{array}{lcl}
\specA\at{\frac{\pi}{4n}}{U}_n\at\phase
&\text{for}&\abs\phase\leq{n}-\tfrac{1}{2},%
\\%
0&\text{for}&\abs\phase\geq{n}+\tfrac{1}{2},%
\end{array}\right.
\end{align*}
and
\begin{align*}
0\leq\at{\calA{U}_n}\at\phase
\leq{{U}}_n\at{n-\tfrac{1}{2}}=
\tfrac{\sqrt{2\ga}}{\sqrt{n}}
\cos\at{\frac{\pi}{2}\at{1-\tfrac{1}{2n}}}=\DO{n^{-3/2}}
\quad\text{for $\babs{\abs\phase-{n}}
\leq\tfrac{1}{2}$}.
\end{align*}
Moreover, we have
\begin{align*}
\calP_\harm\at{{U}_n}\geq\beta\ga\,\specA\at{\frac{\pi}{4n}}^2
\int\limits_{-1+\tfrac{1}{2n}}^{1-\tfrac{1}{2n}}
\cos\at{\frac{\pi}{2}\phase}^2\dint\phase
%=\beta\gamma\at{1-\nDO{n^{-2}}}^2\at{1-\nDO{n^{-3}}}
\geq\beta\gamma\at{1-\nDO{n^{-2}}},
\end{align*}
and this implies $\calP_\harm\at{{U}_n}\to\beta\gamma$ as
$n\to\infty$. Finally, due to Lemma \ref{AP:Lem.AProps} we find
\begin{align*}
\calP_\harm\at{{W}}
=%
\tfrac{1}{2}{\beta}
\int\limits_\Rset\abs{\at{\calA{W}}\at\phase}^2\dint\phase
=%
{\beta}\tfrac{1}{2}\norm{\calA{W}}_2^2
\leq%
\tfrac{1}{2}{\beta}\norm{{W}}_2^2,
\end{align*}
for all ${W}\in\calB_\ga$, and the proof is complete.
\end{proof}
\begin{corollary}
We have
\begin{math}
P\at{\gamma}\geq\beta\gamma=P_\harm\at\ga
\end{math}
for all $\gamma\geq0$.
\end{corollary}
\begin{proof}
Since the case $\beta=0$ is trivial we suppose $\beta>0$. Some
elementary analysis shows that the sequence $\at{{U}_n}_n$ from
\eqref{AP:Def.Harm.MaxSequ} satisfies
\begin{align*}
\calP\at{{U}_n}
\approx&%
\int\limits_{\abs\phase\leq{n}-\tfrac{1}{2}}
\Phi\at{\specA\at{\frac{\pi}{4n}}{U}_n\at\phase}\dint\phase%
\approx%
\tfrac{1}{2}\int\limits_{\abs\phase\leq{n}-\tfrac{1}{2}}{{U}_n\at\phase}^2
\at{{\beta}+\Do{1/\sqrt{n}}}\dint\phase
\approx%
{\beta}\ga,
\end{align*}
where all approximation errors tend to $0$ as $n\to\infty$.
\end{proof}
As already mentioned in the introduction, solitons are genuinely
nonlinear phenomena. In particular, the harmonic chain does not
allow for solitons, and thus the supremum in \eqref{AP:Def.Harm.Sup}
cannot be attained.

\begin{remark} For $\beta>0$ there is no maximiser for $\calP_{\harm}$.
\end{remark}
\begin{proof}
Suppose for contradiction the existence of a maximizer $W_{\max}$
for $\calP_\harm$ in $\calB_\ga$. Then the Lagrangian multiplier
rule implies that there exists a (non-negative) Lagrangian
multiplier $\si^2$ such that
\begin{align*}
\si^2W_{\max}=\partial\calP\ato{W_{\max}}=\beta\widebar{\calA}^2W_{\max},
\end{align*}
i.e., $W_{\max}$ is an $\fspaceL^2$-eigenfunction to
$\widebar{\calA}^2$ with corresponding eigenvalue $\si^2/\beta$.
This is the desired contradiction because the point spectrum of
$\widebar{\calA}^2$ is empty due to Lemma \ref{AP:Lem.AProps}.
\end{proof}
%
%
%--------------------------------------------------------------------
\paragraph*{On the weak convergence of
unimodal, even and non-negative functions in $\fspaceL^2\at{\Rset}$}
%--------------------------------------------------------------------
%
%
For the sake of clarity we proceed with some remarks on the strong
compactness of weak convergent sequences from
$\calU\cap\calN\cap{}\fspaceL^2$ as this problem becomes relevant in
our existence proof for solitons. Strong compactness criterions are,
in principle, provided by the concentration compactness method from
\cite{Lio84}, but since here we consider only functions from
$\calU\cap\calN$ the arguments simplify a lot.
\bigpar%
Consider a sequence
$\at{{W}_n}_n\subset\calS_\ga=\calU\cap\calN\cap{}\calB_\ga$ that
converges weakly in $\fspaceL^2$ to some limit
${W}_\infty\in\calS_\ga$. Passing to a subsequence we can always
assume that
$\ga_n=\tfrac{1}{2}\norm{{W}_n}_2^2\to\ol{\gamma}_\infty$ for some
$\bar{\ga}_\infty$ with $\ga_\infty\leq\ol{\gamma}_\infty\leq\ga$,
where $\ga_\infty=\tfrac{1}{2}\norm{{W}_\infty}_2^2$. For
$\gamma_\infty=\ol{\gamma}_\infty$ the convergence of norms enforces
the convergence ${W}_n\to{{W}}_\infty$ to be strong in $\fspaceL^2$,
and we are done in this case.
\par
\begin{figure}[!ht]%
  \centering{
  \includegraphics[width=0.75\textwidth, draft=\figdraft]%
  {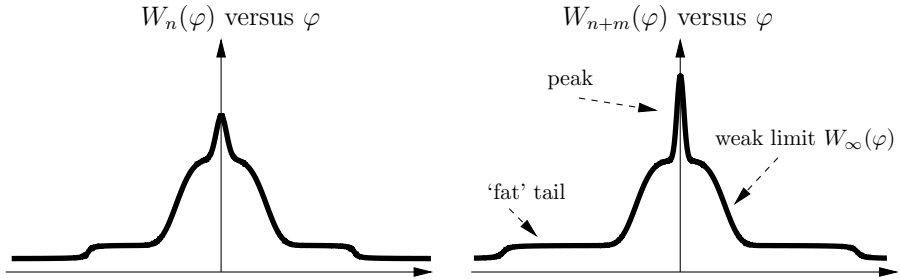}%
  }%
  \caption{%
    On the weak convergence in $\calU\cap\calN\cap{\calB}_\ga$.
    A weak convergent sequence may fail to converge strongly due to the
    formation of a peak at the origin, and/or a `fat' tail.
      }%
  \label{Fig:WeakConv}%
\end{figure}

In the case $\ol{\gamma}_\infty>\gamma_\infty$ the convergence
cannot be strong as some amount of the 'mass' of the measures
$\mu_n={W}_n\at\phase^2\dint\phase$ disappears when passing to
$\mu_\infty={W}_\infty\at\phase^2\dint\phase$. However, since all
functions are non-negative, unimodal, and even, the annihilation of
mass is governed by only two elementary processes, compare Figure
\ref{Fig:WeakConv}. The weakly convergent sequence can form a
\emph{peak at the origin}, and/or a \emph{`fat' tail}, where `fat'
means
\begin{align*}
\lim\limits_{L\to\infty}\lim\limits_{n\to\infty}
\int\limits_{-\infty}^{-L}{W}_n^2\at\phase\dint\phase+
\int\limits_{L}^\infty{W}_n^2\at\phase\dint\phase>0,
\end{align*}
so that some non-negligible amount of the norm is transferred to
infinity.
\par
The first observation is that a peak does not contribute to the
potential energy $\calP$. In fact, if the height of the peak is of
order $1/\eps$ with $\eps\ll1$, then the norm constraint
$\tfrac{1}{2}\norm{{W}}_2\le\ga$ implies that the width of the peak
is of order $\eps^2$, and thus the peak disappears after applying
the averaging operator $\calA$. More rigorously, Lemma
\ref{AP:Lem.AProps} guarantees $\calA{W}_n$ to converge strongly to
$\calA{W}_\infty$ on each compact subset of $\Rset$. Consequently,
if the strong convergence fails due to the formation of a peak only,
we still have $\calP\at{{W}_n}\to\calP\at{{W}_\infty}$.
\par
The formation of a tail, however, is much more crucial as this in
general implies $\calP\at{{W}_n}\nrightarrow\calP\at{{W}_\infty}$.
Remember the maximising sequence ${U}_n$ for $\calP_\harm$ from
\eqref{AP:Def.Harm.MaxSequ} having the property that all mass of the
measures ${U}_n\at\phase^2\dint\phase$ is contained in a fat `tail'
with increasing support and decreasing height, so that the weak
limit is zero. Even worse, since the spectrum of
$\calA^2=\beta^{-1}\partial\calP_\harm$ is continuous, \emph{each}
maximising sequence for $\calP_\harm$ is expected to have this
property, and we conclude that the formation of tails is directly
related to the non-existence of solitons for the harmonic chain.
\par
Our strategy to prove the existence of solitons for nonlinear
potentials is  to show that each sequence, that maximizes $ \calP$
in $\calS_\ga$, is \emph{localised}, so a `fat' tail cannot be
formed. To this end we restrict our considerations to the case of
\emph{super-quadratic} growth, and derive suitable tightness results
under this assumptions.
%
%
%--------------------------------------------------------------------
\subsection{%
\texorpdfstring%
{Existence of solitons for genuinely super-quadratic $\calP$}
{Existence of solitons for genuinely super-quadratic P}
}
\label{sec:ES:ExistProof}%
%--------------------------------------------------------------------
%
For the remainder of this section we require super-quadratic growth
conditions. We start with the assumptions concerning the energy
functional $\calP$ as they appear naturally in our existence proof.
Below in \S\ref{sec:ES:SQCrit} and \S\ref{sec:ES:GSQCrit} we then
discuss the corresponding properties of the atomic interaction
potential $\Phi$.
\begin{definition}
\label{ES:SQ.Def} %
The functional $\calP$ is called \emph{super-quadratic} on
$\calS_\ga$ with $\ga>0$ if
\begin{align}
\label{ES:SQ.Def.Eqn1} %
\calP\at{s{W}}\geq{s^2}\calP\at{{W}}
\qquad\text{for}\;\text{all}\;\;\;%
{W}\in\calS_\ga
\;\;\;\text{and}\;\;\;%
1\leq{s}\leq\frac{\sqrt{2\ga}}{\norm{{W}}_2}.
\end{align}
Moreover, $\calP$ is called \emph{genuinely super-quadratic} on
$\calS_\ga$ if in addition
\begin{align*}
P\at{\gamma}>P_\harm\at{\gamma}={\beta}\gamma.
\end{align*}
\end{definition}
Notice that for each $\ga>0$ the harmonic functional $\calP_\harm$
is super-quadratic, but not genuinely super-quadratic in
$\calS_\ga$.
\begin{remark}
Let $\calP$ be super-quadratic on $\calS_{\ga}$. Then it is
super-quadratic on $\calS_{\tilde{\ga}}$ for all
$0\leq\tilde\ga\leq\gamma$, and we have
\begin{align*}
\frac{P\at{\gamma_2}}{\gamma_2}
\geq%
\frac{P\at{\gamma_1}}{\gamma_1}\geq{\beta}
\end{align*}
for all $0\leq\gamma_1\leq\gamma_2\leq\gamma$. In particular, if
$\calP$ is genuinely super-quadratic for $\gamma_1\leq\gamma$, so it
is for every $\gamma_2$ with $\gamma_1\leq\gamma_2\leq\gamma$.
\end{remark}
\begin{remark}
Let $\calP$ be a super-quadratic on $\calS_{\ga}$. Then, we have
\begin{align}
\label{ES:QuadraticEstimate}%
\skp{\partial\calP\ato{{W}}}{{W}}\geq
2\calP\at{{W}}
\end{align}
for all ${W}\in\calS_{\ga}$.
\end{remark}
\begin{proof}
In view of the continuity properties of $\calP$ and $\partial\calP$
it is sufficient to consider the case $\norm{{W}}_2<\ga$. For
sufficiently small $\eps$ \eqref{ES:SQ.Def.Eqn1} implies
\begin{align*}
\eps^{-1}\at{\calP\at{\at{1+\eps}{W}}-\calP\at{{W}}}
\geq
\eps^{-1}\bat{\at{1+\eps^2}-1}\calP\at{W}=\at{2+\eps}\calP\at{W}
\end{align*}
and the limit $\eps\to0$ gives \eqref{ES:QuadraticEstimate}.
\end{proof}
We now formulate or main technical result concerning the tightness
of maximising sequences. Roughly spoken, `fat' tails are not
energetically optimal as their contributions to $\calP$ and
$\calP_\harm$ are comparable, and peaks are not optimal as they do
not contribute to the potential energy at all. These naive
explanations can be stated rigorously as follows.
\begin{lemma}
\label{ES:Tightness.Lemma1} %
For any $\delta>{0}$ the set
\begin{align*}
\calS_{\ga,\,\delta}=\left\{{W}\in\calS_\ga\;:\;
\calP\at{{W}}-\beta\tfrac{1}{2}\norm{{W}}_2^2\geq\delta\right\}
\end{align*}
is closed under weak convergence.
\end{lemma}
\begin{proof}
Let $\at{{W}_n}_n\subset\calS_\ga$ be a given sequence such that
${W}_n\to{{W}_\infty}$ as $n\to\infty$ weakly in $\fspaceL^2$, and
$\calP\at{{W}_n}\geq\beta\ga_n+\delta$ with
$\ga_n=\tfrac{1}{2}\norm{{W}_n}_2^2$. Without loss of generality we
can assume that $\tfrac{1}{2}\norm{{W}_n}_2^2\to\ol{\gamma}_\infty$
for some $\ol{\gamma}_\infty$ with
$\gamma_\infty\leq\ol{\gamma}_\infty\leq{\ga}$ and
$\ga_\infty=\tfrac{1}{2}\norm{{W}_\infty}_2^2$. It remains to show
that $\calP\at{{W}_\infty}\geq\beta\ga_\infty+\delta$. For
$n\in\Nset\cup\{\infty\}$ we set
\begin{align}
\label{ES:Tightness.Lemma1.Proof2} %
\widetilde{{W}}_n\deq{W}_n|_{\ccinterval{-m}{+m}},\qquad
\widehat{{W}}_n\deq{W}_n-\widetilde{{W}}_n,
\end{align}
where $m>0$ is some constant to be chosen below, and this definition
implies
\begin{align}
\label{ES:Tightness.Lemma1.Proof12} %
\norm{{W}_n}_2^2
=%
\norm{\widetilde{{W}}_n}_2^2+\norm{\widehat{{W}}_n}_2^2.
\end{align}
Our strategy for this proof is to establish the approximations
\begin{align*}
\calP\at{{W}_n}
\approx%
\calP\nat{\widetilde{{W}}_n}+\calP\nat{\widehat{{W}}_n},\quad
\calP\nat{\widehat{{W}}_n}
\approx%
\calP_\harm\nat{\widehat{{W}}_n},
\end{align*}
where the approximation error becomes arbitrary small if both $m$
and $n$ are sufficiently large. To show this we fix $\eps>0$, and
suppose $m$ to be sufficiently large such that
\begin{align}
\label{ES:Tightness.Lemma1.Proof1} %
\tfrac{1}{2}\norm{\widehat{{W}}_\infty}^2_2\leq\eps,
\qquad%
\abs{\calP\at{{W}_\infty}-\calP\nat{\widetilde{{W}}_\infty}}\leq\eps
,\qquad%
\int\limits_{m-1}^m{{W}_\infty}\at{\phase}\dint{\phase}
\leq\eps.
\end{align}
Such a choice for $m$ exists as
$\widetilde{{W}}_\infty\to{{W}}_\infty$ strongly in
$\fspaceL^2$ as $m\to\infty$.
Since $m$ is finite Lemma \ref{AP:Lem.AProps} provides
$\calA\widetilde{{W}}_n\to\calA\widetilde{{W}}_\infty$ strongly in
$\fspaceL^2$ as $n\to\infty$, and thus we find
\begin{align}
\label{ES:Tightness.Lemma1.Proof3}
\abs{\norm{\widetilde{{W}}_\infty}_2^2-
\norm{\widetilde{{W}}_n}_2^2}\leq\eps
,\qquad%
\abs{\calP\nat{\widetilde{{W}}_\infty}-
\calP\nat{\widetilde{{W}}_n}}
\leq%
\eps,\qquad
\abs{ \int\limits_{m-1}^m{W}_\infty\at{\phase}-
W_n\at{\phase}\dint{\phase}}\leq\eps
\end{align}
for all sufficiently  large $n$. Moreover, combining
\eqref{ES:Tightness.Lemma1.Proof1}$_1$ and
\eqref{ES:Tightness.Lemma1.Proof3}$_1$ with
\eqref{ES:Tightness.Lemma1.Proof12} and $\ga_n\to\ol{\ga}_\infty$
shows that
\begin{align}
\label{ES:Tightness.Lemma1.Proof6}
\tfrac{1}{2}\norm{\widehat{{W}}_{n}}_2^2=
\gamma_n-\tfrac{1}{2}\norm{\widetilde{{W}}_{n}}_2^2 \leq
\ol{\gamma}_\infty-
\tfrac{1}{2}\norm{\widetilde{{W}}_{\infty}}_2^2+C\eps\leq
\ol{\gamma}_\infty-\gamma_\infty+C\eps
\end{align}
with $C$ independent of $n$ and $\eps$. In virtue of
\eqref{ES:Tightness.Lemma1.Proof1}$_3$ and
\eqref{ES:Tightness.Lemma1.Proof3}$_3$, and since all functions
$W_n$ are unimodal, non-negative, and even,
we also obtain %
\begin{align*}
0\leq{\nat{\calA\widehat{W}_{n}}\at{\phase}}
\leq{\at{\calA{W}_{n}}\at{\phase}}
\leq%
{\at{\calA{W}_{n}}\at{m-\tfrac{1}{2}}}
\leq%
\int\limits_{m-1}^{m}{W}_{n}\at{\tilde\phase}\dint\tilde{\phase}
\leq%
2\eps
\end{align*}
for large $n$ and all $\phase$ with
$\abs{\phase}\geq{m}-\tfrac{1}{2}$. This provides
\begin{align}
\label{ES:Tightness.Lemma1.Proof4}
\norm{\calA\widehat{{W}}_{n}}_\infty\leq{C}\eps,
\qquad
\int\limits_{\babs{\abs{\phase}-m}
\leq%
\tfrac{1}{2}}\Phi\at{\calA{W}_{n}}\dint\phase\leq{C}\eps,
\end{align}
and exploiting the expansions of $\Phi\at{r}$ for small $r$, compare
\eqref{AP:PotAss.Eqn1}, we find
\begin{align}
\label{ES:Tightness.Lemma1.Proof5}
\calP\nat{\widehat{{W}}_{n}}&\leq\tfrac{1}{2}\at{\beta+\Do{\eps}}
\int\limits_{\Rset}\nat{\calA\widehat{{W}}_{n}}\at{\phase}^2\dint\phase
\leq\tfrac{1}{2}\at{{\beta}+\Do{\eps}}\norm{\widehat{{W}}_{n}}_2^2
\leq{\beta}\at{\ol{\gamma}_\infty-\gamma_\infty}+ C\eps
\end{align}
thanks to \eqref{ES:Tightness.Lemma1.Proof6}. Finally, due to
\eqref{ES:Tightness.Lemma1.Proof2} we have
\begin{align*}
\calP\at{{W}_{n}}&=
\int\limits_{\abs{\phase}\leq{m}-\tfrac{1}{2}}
\Phi\at{\calA{W}_{n}}\dint\phase
+\int\limits_{\babs{\abs{\phase}-m}\leq\tfrac{1}{2}}
\Phi\at{\calA{W}_{n}}\dint\phase
+\int\limits_{\abs{\phase}\geq{m}+\tfrac{1}{2}}
\Phi\at{\calA{W}_{n}}\dint\phase
\\&=%
\int\limits_{\abs{\phase}\leq{m}-\tfrac{1}{2}}
\Phi\nat{\calA\widetilde{{W}}_{n}}\dint\phase
+\int\limits_{\babs{\abs{\phase}-m}\leq\tfrac{1}{2}}
\Phi\at{\calA{W}_{n}}\dint\phase
+\int\limits_{\abs{\phase}\geq{m}+\tfrac{1}{2}}
\Phi\nat{\calA\widehat{{W}}_{n}}\dint\phase
\\&\leq
\calP\nat{\widetilde{{W}}_{n}}
+\int\limits_{\babs{\abs{\phase}-m}\leq\tfrac{1}{2}}
\Phi\nat{\calA{W}_{n}}\dint\phase+
\calP\nat{\widehat{{W}}_{n}},
\end{align*}
and \eqref{ES:Tightness.Lemma1.Proof1}$_2$ and
\eqref{ES:Tightness.Lemma1.Proof3}$_2$ combined with
\eqref{ES:Tightness.Lemma1.Proof4} and
\eqref{ES:Tightness.Lemma1.Proof5} imply
\begin{align*}
\calP\at{{W}_{n}}\leq%
\calP\nat{\widetilde{{W}}_{\infty}}+\calP\nat{\widehat{{W}}_{n}}+C\eps
\leq%
\calP\nat{{W}_{\infty}}+
{\beta}\at{\ol{\gamma}_\infty-\gamma_\infty}+C\eps.
\end{align*}
Using the assumption we conclude that
$\calP\nat{{W}_{\infty}}\geq\ga_\infty+\,\delta-C\eps$, and this
completes the proof because $\eps$ was chosen arbitrarily.
\end{proof}
As a direct consequence of Lemma \ref{ES:Tightness.Lemma1} we find
in the genuinely super-quadratic case that each maximising sequence
must be localized and contains a strongly convergent subsequence.

\begin{corollary}
\label{ES:Tightness.Corr2} %
Let $\calP$ be genuinely super-quadratic on $\calS_\gamma$, and
suppose that the sequence $\at{{W}_n}_n\subset\calS_\ga$ is a
maximising sequence for $\calP$ on $S_\ga$, that means
\begin{align*}
\lim_{n\to\infty}\calP\at{{W}_n}
=%
P\at{\gamma}={\beta}\ga+\delta>P_\harm\at{\ga},
\end{align*}
for some $\delta>0$. Then, there exists a subsequence, still denoted
by $\at{{W}_n}_n$, and ${W}_\infty\in\partial\calS_\ga$ such that
${W}_n\to{{W}_\infty}$ strongly in $\fspaceL^2$, and hence
$\calP\at{{W}_\infty}=P\at\gamma$.
\end{corollary}
\begin{proof}
We choose the subsequence and ${W}_\infty\in\calS_\ga$ such that
${W}_n\to{{W}_\infty}$ weakly in $\fspaceL^2$. Thanks to
\eqref{ES:SQ.Def.Eqn1} we know that
$\calP\at{s{W}_n}\geq{}s^2\calP\at{{W}_n}$ for all $s>1$, and this
implies $\ga_n=\tfrac{1}{2}\norm{{W}_n}_2^2\to\ga$ as $n\to\infty$,
because otherwise the sequence $\at{{W}_n}_n$ could not be
maximising. Therefore, with
\begin{align*}
\widetilde{{W}}_n=
\frac{\sqrt{2\ga}}{\sqrt{2\ga_n}}{W}_n\in\partial\calS_\ga
\end{align*}
we find $\calP\nat{\widetilde{{W}}_n}\to\beta\gamma+\delta$ and
$\widetilde{{W}}_n\to{W}_\infty$ weakly in $\calS_\ga$, and Lemma
\ref{ES:Tightness.Lemma1} provides
\begin{align*}
\calP\at{{W}_\infty}\geq\beta\ga_\infty+\delta
\end{align*}
with $\gamma_\infty=\tfrac{1}{2}\norm{{W}_\infty}_2^2\leq\ga$. In
order to show ${W}_\infty\in\partial\calS_\ga$ we use again
\eqref{ES:SQ.Def.Eqn1} to obtain
\begin{align*}
\beta\gamma+\delta=P\at\ga\geq
\calP\at{\sqrt{\tfrac{\ga}{\ga_\infty}}{W}_\infty}
\geq%
\tfrac{\ga}{\ga_\infty}
\calP\at{{W}_\infty}
\geq%
\tfrac{\ga}{\ga_\infty}\at{\beta\ga_\infty+\delta}
\geq\beta{\ga}+\tfrac{\ga}{\ga_\infty}\delta.
\end{align*}
Since $\delta>0$ we conclude that $\gamma_\infty=\gamma$, that means
$\norm{W_n}_2\to\norm{W_\infty}_2$, and this implies that the
convergence ${W}_n\to{W}_\infty$ is strong in $\fspaceL^2$.
\end{proof}
The combination of Theorem \ref{IS:Cone2.Theo} and Corollary
\ref{ES:Tightness.Corr2} immediately provides the desired existence
result for non-negative and unimodal solitons.
\begin{corollary}
\label{ES:Corr.Existence} %
If $\calP$ is genuinely super-quadratic on $\calS_\gamma$, then
there exists a maximiser ${W}$ of $\calP$ in $\calS_\ga$. This
maximiser is a non-negative and unimodal soliton with
$\tfrac{1}{2}\norm{{W}}^2_2=\ga$.
\end{corollary}
Finally, we characterise the soliton speed of maximizers of $\calP$,
and mention that an analogous result holds for wave trains (provided
that $\calP$ is super-quadratic).
\begin{remark} %
The soliton from Corollary \ref{ES:Corr.Existence} is super-sonic,
that means $\sigma^2>\beta$.
\end{remark}
\begin{proof}
Testing the soliton equation \eqref{OV:TWEqn} with ${W}$, and using
\eqref{ES:QuadraticEstimate}, we find
$\si^2\norm{{W}}_2^2\geq2\calP\at{{W}}>2\beta\ga$.
\end{proof}
%
%--------------------------------------------------------------------
\subsection{%
\texorpdfstring%
{Criterions for super-quadratic $\calP$}%
{Criterions for super-quadratic P}%
}%
\label{sec:ES:SQCrit}
%--------------------------------------------------------------------
%
%
\begin{definition}
\label{SQ:Def.1} %
The potential $\Phi$ is called super-quadratic on the interval
$\ccinterval{0}{\sqrt{2\ga}}$ if
\begin{align*}
\Phi\at{sr}\geq{s^2}\Phi\at{r}
\end{align*}
holds for all $r\geq0$ and all $s\geq1$ with
$rs\in\ccinterval{0}{\sqrt{2\ga}}$.
\end{definition}
\begin{remark}
If the potential $\Phi$ is super-quadratic on the
interval $\ccinterval{0}{\sqrt{2\ga}}$, then $\calP$ is
super-quadratic on $\calS_\ga$.
\end{remark}
\begin{proof}
Let ${W}\in\calS_\ga$ be fixed, and  $s$ be arbitrary with
$1\leq{s}\leq{\sqrt{2\ga}}/\norm{{W}}_2$. For all
$r=\at{\calA{W}}\at{\phase}\leq\sqrt{\norm{{W}}_2}$ we have
$rs\leq\sqrt{2\ga}$, and hence
$\Phi\at{s\at{\calA{W}}\at{\phase}}\geq{s^2}\Phi\at{\at{\calA{W}}\at{\phase}}$,
for (almost) all $\phase\in\Rset$. Finally, integration w.r.t.
$\phase$ yields the desired result.
\end{proof}
Definition \ref{SQ:Def.1} implies that $\Phi$ is super-quadratic on
the interval $\ccinterval{0}{\sqrt{2\ga}}$ if and only if the
function
\begin{align}
\label{SQ:Def.BetaFunct}
\beta_{\Phi}\at{r}\deq{2}\frac{\Phi\at{r}}{r^2}
\end{align}
is non-decreasing on $\ccinterval{0}{\sqrt{2\ga}}$ (notice that
$\beta_{\Phi}\at{0}=\beta$ in the sense of Assumption
\ref{AP:PotAss}).  To obtain further characterisations of
super-quadratic growth we consider the following differential
inequalities.
\begin{enumerate}
\item[$\at{\text{C1}}$]%
$\Phi^\prime\at{r}r-2\Phi\at{r}\geq0$ for all
$r\in\ccinterval{0}{\sqrt{2\ga}}$,
\item[$\at{\text{C2}}$]%
$\Phi^{\prime\prime}\at{r}r-\Phi^\prime\at{r}\geq0$ for all
$r\in\ccinterval{0}{\sqrt{2\ga}}$,
\item[$\at{\text{C3}}$]%
$\Phi^{\prime\prime\prime}\at{r}\geq0$ for all
$r\in\ccinterval{0}{\sqrt{2\ga}}$.
\end{enumerate}
\begin{remark}
%\label{SQ:Remark.Crit1}
%
For all $\ga>0$ and all sufficiently smooth potentials $\Phi$ with
$\Phi\at{0}=\Phi^\prime\at{0}=0$ we have
\begin{align*}
\at{\text{C3}}\quad\Longrightarrow\quad
\at{\text{C2}}\quad\Longrightarrow\quad%
\at{\text{C1}},
\end{align*}
where $\at{\text{C1}}$ holds if and only if $\Phi$ is
super-quadratic on $\ccinterval{0}{\sqrt{2\ga}}$.
\end{remark}
\begin{proof}
$\at{\text{C3}}$ is equivalent to
$\frac{\dint}{\dint{r}}\at{\Phi^{\prime\prime}\at{r}r-\Phi^\prime\at{r}}\geq0$,
and in view of $\Phi^{\prime\prime}\at{0}0-\Phi^\prime\at{0}=0$ we
conclude that $\at{\text{C3}}\Rightarrow\at{\text{C2}}$. Moreover,
the implication $\at{\text{C2}}\Rightarrow\at{\text{C1}}$  can be
proven similarly, and $\at{\text{C1}}$ is equivalent to
$\frac{\dint}{\dint{r}}\beta_{\Phi}\at{r}\geq0$.
\end{proof}
We proceed with a remark concerning the relation between
super-quadratic growth and the convexity of $\Phi$. In our context,
of course, super-quadratic potentials being not convex are
forbidden, but solitons can still be shown to exist for such
potentials, see \cite{FW94} and our comments below.
\begin{remark}
\label{SQ:Rem.Convexity.Pot}
$\at{\text{C2}}$ implies the convexity of $\Phi$ (on the interval
$\ccinterval{0}{\sqrt{2\ga}}$), but there exists non-convex
potentials satisfying $\at{\text{C1}}$
\end{remark}
\begin{proof}
Suppose that $\Phi$ satisfies $\at{\text{C2}}$. Then, the comparison
principle for ODEs gives $\Phi^\prime\at{r}\geq\beta{r}$ for some
$\beta\geq0$, and $\at{\text{C2}}$ implies
$\Phi^{\prime\prime}\at{r}\geq\beta$ for all $r>0$. Now let $\eta>0$
be arbitrary, and consider the potential
\begin{align*}
\Phi_\eta\at{r}=r^2\at{1+2\pi^{-1}\arctan\at{\eta\at{r-1}}},
\end{align*} %
which is super-quadratic on $\cointerval{0}{\infty}$ as the function
$\beta_{\Phi_\eta}$ is strictly increasing by construction. A direct
calculation yields
\begin{align*}
\Phi^{\prime\prime}_\eta\at{1+\eta^{-1}}=
\pi^{-1}\at{3 + 3\pi + 2\eta - \eta^2},
\end{align*} %
hence $\Phi_\eta$ is not convex for large $\eta$.
\end{proof}
Another remark concerns the convexity of forces, which become
important in the context of atomistic Riemann problems in FPU
chains, compare  \cite{HR08a}.
\begin{remark}
%\label{SQ:Rem.Convexity.Forces}
$\at{\text{C3}}$ implies the convexity of $\Phi^\prime$ (on the
interval $\ccinterval{0}{\sqrt{2\ga}}$), but there exists potentials
$\Phi$ that satisfy $\at{\text{C2}}$ with non-convex derivative.
\end{remark}
\begin{proof}
The first statement is obvious, and towards the second claim we
argue as follows. We choose a non-negative, but not monotonically
increasing function $h$ with $h\at{0}=0$, and compute $\Phi^\prime$
as solution to the ODE
$h\at{r}=\Phi^{\prime\prime}\at{r}r-\Phi^\prime\at{r}$. Then, each
local extremum of $h$ for $r>0$ is a turning point of $\Phi^\prime$,
and vice versa.
\end{proof}
%
%
%--------------------------------------------------------------------
\subsection{%
\texorpdfstring%
{Criterions for genuinely super-quadratic $\calP$}%
{Criterions for genuinely super-quadratic P}%
}%
\label{sec:ES:GSQCrit}
%--------------------------------------------------------------------
%
In order to complete the existence proof for solitons we must show
that for a given super-quadratic potential $\Phi$ the corresponding
energy functional $\calP$ is in fact genuinely super-quadratic on
$\calS_\ga$. In the simplest case there is no harmonic contribution
to $\calP$ at all, and then there exist solitons with arbitrary
small $\ga$. This holds in particular  for all homogenous potentials
$\Phi\at{r}=cr^\alpha$ with $c>0$ and $\alpha>2$.
\begin{remark}
Let $\Phi$ be super-quadratic on $\cointerval{0}{\infty}$ with
$\beta=\Phi^{\prime\prime}\at{0}=0$ and $\Phi\at{r}>0$ for all
$r>0$. Then, $\calP$ is genuinely super-quadratic on $\calS_\ga$ for
all $\gamma>0$.
\end{remark}
\par
The case $\beta>0$ is more involved and needs a better understanding
of the balance between the harmonic and non-harmonic contributions
to $\calP$. Our strategy in this case is to find a particular
function ${W}_0$ such that
$\calP\at{{W}_0}>\tfrac{1}{2}\beta\norm{{W}_0}_2^2$, and this in
turn implies the existence of solitons for all
$\ga\geq\tfrac{1}{2}\norm{{W}_0}_2^2$.
\begin{lemma}
\label{SQ:Lemma.Crit1}%
Suppose that $\Phi$ is super-quadratic on $\cointerval{0}{\infty}$
and that the function $\beta_\Phi$ from \eqref{SQ:Def.BetaFunct}
satisfies
\begin{math}
\lim_{r\to\infty}\beta_\phi\at{r}
>%
\tfrac{3}{2}\beta
=%
\frac{3}{2}\beta_\Phi\at{0}
\end{math}. %
Then, $\calP$ is genuinely super-quadratic on $\calS_\ga$ for all
sufficiently large $\gamma$.
\end{lemma}
\begin{proof}
For fixed $0<\eps<1$ and ${W}=\sqrt{2\ga}\,{W}_\CL$ with
${W}_\CL$ as in \eqref{HSL:WCL.Def} we find
\begin{align*}
\calP\at{W}
&=%
2\int\limits_0^1\Phi\nat{\sqrt{2\ga}s}\dint{s}\geq%
2\ga\int\limits_0^1\beta_\Phi\nat{\sqrt{2\ga}s}s^2\dint{s}
\\&\geq%
2\ga\beta_\Phi\nat{0}\int\limits_0^{\eps}s^2\dint{s}+
2\ga\beta_\Phi\nat{\eps\sqrt{2\ga}}\int\limits_{\eps}^1s^2\dint{s}
%\\&
=%
\tfrac{2}{3}\beta\gamma\at{\eps^3+
\frac{\beta_\Phi\nat{\sqrt{2\ga}\eps}}{\beta}\at{1-\eps^3}},
\end{align*}
and conclude that $\calP\at{W}>\beta\ga$ for all sufficiently large
$\ga$.
\end{proof}
Lemma \ref{SQ:Lemma.Crit1} implies the existence of solitons for
weakly super-quadratic potentials as for instance
\begin{align*}
\Phi\at{r}=\tfrac{\beta}{2}r^2\at{1+c\ln\at{1+r}},
\qquad
\Phi\at{r}=\tfrac{\beta}{2}r^2\at{1+d\arctan\at{r}}
\end{align*}
with $\beta\geq0$, $c>0$ arbitrary, and $d>0$ sufficiently large.
\bigpar
Next we evaluate the sequence $\at{U_n}_n$ from
\eqref{AP:Def.Harm.MaxSequ} and find an existence criterion for
solitons that is very close to that given in \cite{FW94}.
\begin{lemma}
\label{SQ:Lemma.Crit2}%
Let $\Phi$ be super-quadratic on $\cointerval{0}{\infty}$, and
suppose
\begin{align*}
\Phi\at{r}\geq\tfrac{1}{2}{\beta}r^2+\eps{r}^p
\end{align*}
for all $r\geq0$ with two constants $\eps>0$ and $p>2$. Then there
exists $\gamma_0>0$ such that $\calP$ is genuinely super-quadratic
on $\calS_\ga$ for all $\gamma>\gamma_0$. Moreover, $\beta=0$ or
$2<p<6$ implies $\gamma_0=0$.
\end{lemma}
\begin{proof}
Let $\ga>0$ be arbitrary, and consider the sequence $\at{{U}_n}_n$
from \eqref{AP:Def.Harm.MaxSequ}. Then,
\begin{align*}
\calP\at{{U}_n}-\calP_\harm\at{{U}_n}\geq
\eps\calP_\nl\at{{U}_n}
\end{align*}
with
\begin{align*}
\calP_\nl\at{{U}_n}&=
\int\limits_{\Rset}\at{\calA{U}_n}^p\dint\phase
\geq\int\limits_{\abs{\phase}\leq{n}-
\tfrac{1}{2}}\at{\specA\at{\tfrac{\pi}{4n}}{U}_n}^p\dint\phase
\\&\geq%
\at{\specA\at{\tfrac{\pi}{4n}}}^p\at{\tfrac{2{\ga}}{n}}^{p/2}n
\int\limits_{-1+\tfrac{1}{2n}}^{1-\tfrac{1}{2n}}
\at{\cos\at{\tfrac{\pi}{2}\phase}}^p\dint\phase \geq{c}n^{1-p/2} .
\end{align*}
We conclude
\begin{align*}
\calP\at{{U}_n}
\geq%
\calP_\harm\at{{U}_n}+c{\gamma}^{p/2}n^{1-p/2}>0
\end{align*}
for some positive constant $c>0$, and according to Lemma
\ref{AP:Lemma.MaxSequ} there exists a constant $\tilde{c}>0$ such
that
\begin{align*}
\calP\at{{U}_n}\geq{c}{\gamma}^{p/2}n^{1-p/2}+
\beta{\gamma}\at{1-\tilde{c}n^{-2}}.
\end{align*}
Finally, for $\beta=0$, or ${\gamma}$  sufficiently large, or
$-1+p/2<2$ and $n$ large,  we find $\calP\at{{U}_n}>\beta\gamma$.
\end{proof}
As an application of Lemma \ref{SQ:Lemma.Crit2} we find the
following existence result for unimodal solitons with non-negative
$W$: Let $\Phi\at{r}$ be analytic with non-negative coefficients,
i.e.,
$\Phi\at{r}=\tfrac{1}{2}{\beta}r^2+\sum_{i=3}^\infty\kappa_ir^i$
with $\kappa_i\geq0$ for all $i\geq3$. Then $\Phi$ is
super-quadratic on $\cointerval{0}{\infty}$ and genuinely
super-quadratic for large $\ga$. Moreover, if at least one of the
coefficients $\kappa_3$, $\kappa_4$, and $\kappa_5$ is positive,
then $\Phi$ is genuinely super-quadratic for all $\gamma>0$.
\par
Moreover, since the traveling wave equation \eqref{OV:TWEqn} is
invariant under the reflection symmetry
\begin{align*}
W\rightsquigarrow-W,\quad\Phi\at{r}\rightsquigarrow\Phi\at{-r},
\end{align*}
we find also existence results for solitons with non-positive $W$.
For instance, the Toda potential $\Phi_{\toda}\at{r}=\mhexp{-r}+r-1$
is not super-quadratic for $r\geq0$ but the reflected potential
$\widetilde{\Phi}_{\toda}\at{r}=\mhexp{r}-r-1$ has solitons with
arbitrary small $\ga$, compare Lemma \ref{SQ:Lemma.Crit2}. From this
we infer that the Toda chain  allows for solitons with $W\leq0$.

\bigpar%
Finally, we summarise some other super-quadratic growth conditions
for $\Phi$ under which the existence of solitons was proved by other
authors.
\begin{enumerate}
\item
$\Phi^\prime\at{r}r>2\Phi\at{r}$: Friesecke and Wattis \cite{FW94}
prove the existence of super-sonic solitons with prescribed
potential energy $\calP\geq{P}_0$ above some critical energy
$P_0\geq0$. Moreover,
$\Phi\at{r}=\tfrac{1}{2}{\beta}r^2+\eps{r}^p\at{1+\Do{{r}}}$ with
$\eps>0$ and $p$ as in Lemma \ref{SQ:Lemma.Crit2} implies $P_0=0$.
\item
$\Phi\at{r}=\tfrac{\beta}{2}r^2+\Phi_\nl\at{r}$ and
$\Phi_\nl^\prime\at{r}r\geq\alpha\Phi_\nl\at{r}$ for all $r\geq0$
and some $\alpha>2$: Smets and Willem \cite{SW97} establish the
existence of solitons with prescribed super-sonic speed
$\si^2>\beta=\Phi^{\prime\prime}\at{0}$
\item
$\Phi_\nl^\prime\at{r}r\geq\al\Phi_\nl\at{r}$ for all $r\geq0$ and
some $\al>2$, or
$\Phi_\nl^{\prime\prime}\at{r}r\geq\tilde{\al}\Phi_\nl^\prime\at{r}$
for all $r\geq0$ and some $\tilde{\al}>1$: Pankov and Pfl\"uger
\cite{PP00} prove under both assumptions the existence of a family
of solitons parametrised by $\si^2>\beta=\Phi^{\prime\prime}\at{0}$.
\item
$\Phi^\prime\at{r}\geq0$ for all $r\geq0$ and
$\liminf_{r\to\infty}r^{-\al}\at{\Phi^\prime\at{r}r-\alpha\Phi\at{r}}>0$
and some $\alpha>2$: Schwetlick and Zimmer \cite{SZ07} show that for
each super-sonic speed
$\si^2>\si_\crit\geq\beta=\Phi^{\prime\prime}\at{0}$ there exists a
soliton.
\end{enumerate}
All these existence results imply that the soliton profile $W$
belongs to $\calN$, but since they do not require the convexity of
$\Phi$, they do not provide ${W}\in\calU$.
%
%
%
%
%--------------------------------------------------------------------
\subsection%
{Solitons as limits of wave trains}%
\label{sec:ES.GC}%
%--------------------------------------------------------------------
%
It is very natural to ask whether or not wave trains converge to
solitons when the periodicity length $\Len$ tends to $\infty$. In
this section we establish such a convergence result for unimodal and
non-negative wave trains. To this end we allow for arbitrary values
of $\Len\in\ocinterval{0}{\infty}$, and write $\calA_\Len$ for the
operator $\calA$ acting on
$\fspaceL^2\at{\ccinterval{-\Len}{\Len}}$. Consequently, we
introduce
\begin{align*}
\calS_{\Len,\,\ga}:=\calU\cap\calN\cap\calB_{\Len,\,\ga}
\end{align*}
where $\calB_{L,\,\ga}$ denotes the ball of radius $\sqrt{2\ga}$ in
$\fspaceL^2\at{\ccinterval{-\Len}{\Len}}$, and consider
\begin{align*}
P_\Len\deq\sup\limits_{W\in\calS_{\Len,\,\ga}}
\int\limits_{-\Len}^\Len\Phi\bat{\at{\calA_\Len{W}}\at\phase}\dint\phase.
\end{align*}
Moreover, we define an \emph{embedding} operator
$E_\Len:\calS_{\Len,\,\ga}\to\calS_{\infty,\,\ga}$ by
\begin{align*}
\at{E_\Len{W}_\Len}\at\phase=%
\left\{%
\begin{array}{ll}
{W}_\Len\at\phase&\text{for $\abs{\phase}\geq{\Len}$},
\\%
0&\text{else}.
\end{array}\right.
\qquad
\end{align*}
\bigpar
Inspired by the notion of $\Gamma$-convergence we show that the
energy of each periodic profile can be approximated by localised
profiles, and prove that each localised profile can be recovered by
periodic profiles.
\begin{lemma}
\label{GC:Lem.Conv1} %
For each $\ga>0$ and $\Len<\infty$ there exists a constant
$C_{\Len,\,\ga}$ of order $\nDo{\sqrt{\ga/\Len}}$ such that
\begin{align}
\label{GC:Lem.Conv1.Eqn0}
\calP_\infty\at{E_\Len{W}_\Len}+C_{\Len,\,\ga}\geq
\calP_\Len\at{{W}_\Len}\geq\calP_\infty\at{E_\Len{W}_\Len}
\end{align}
holds for all ${W}_\Len\in\calS_{\Len,\,\ga}$. Moreover, for
any ${W}_\infty\in\calS_{\infty,\,\ga}$ there exists a family
of functions  $\at{{W}_\Len}_{\Len<\infty}$ such that
\begin{align}
\label{GC:Lem.Conv1.Eqn1}
\calP_\Len\at{{W}_\Len}%
\xrightarrow{\Len\to\infty}\calP_\infty\at{{W}_\infty}.
\end{align}

\end{lemma}
\begin{proof}
First let ${W}_\Len\in\calS_{\Len,\,\ga}$ be fixed, and
notice that
\begin{align}
\label{GC:Lem.Conv1.Eqn2}
\at{\calA_\infty{}E_\Len{W}_\Len}\at\phase=%
\at{E_\Len\calA_\Len{W}_\Len}\at\phase,\qquad
\babs{\abs{\phase}-\Len}\geq\tfrac{1}{2},%
\end{align}
hold by construction. For
$\babs{\abs{\phase}-\Len}\leq\tfrac{1}{2}$ we have
$0\leq{W}_\Len\at{\phase}\leq{W}_\Len\at{L-1}$ due to
${W}_\Len\in\calU\cap\calN$, and hence
\begin{align*}
0\leq\at{\calA_\infty{}E_\Len{W}_\Len}\at\phase
\leq%
\at{E_\Len\calA_\Len{W}_\Len}\at\phase
\leq%
\int\limits_{\phase-1/2}^{\phase+1/2}%
{W}_\Len\at{\tilde\phase}\dint{\tilde\phase}
=%
{W}_\Len\at{\Len-1}.
\end{align*}
Moreover, from ${W}_\Len\in\calU\cap\calN$ we infer
$\at{\calA_\infty{}E_\Len{W}_\Len}\at\phase\leq{}\at{E_\Len\calA_\Len{W}_\Len}\at\phase$
and
\begin{align*}
2\ga\geq\int\limits_{-\Len+1}^{\Len-1}%
{W}_\Len\at{\phase}^2\dint{\phase}%
\geq2\at{\Len-1}\at{{W}_\Len\at{\Len-1}}^2,
\end{align*}
and therefore
\begin{align}
\label{GC:Lem.Conv1.Eqn3}
0\leq%
\at{\calA_\infty{}E_\Len{W}_\Len}\at\phase
\leq%
\at{E_\Len\calA_\Len{W}_\Len}\at\phase
\leq%
\eps,\qquad
\babs{\abs{\phase}-\Len}\leq\tfrac{1}{2}.%
\end{align}
with $\eps=\sqrt{\ga/\at{\Len-1}}$.  The estimate
\eqref{GC:Lem.Conv1.Eqn0} now follows from \eqref{GC:Lem.Conv1.Eqn2}
and \eqref{GC:Lem.Conv1.Eqn3} via
\begin{align*}
0&\leq\calP_\Len\at{{W}_\Len}-\calP_\infty\at{E_\Len{W}_\Len}
\\&\leq%
\int\limits_{-\infty}^\infty%
\Phi\Bat{\at{E_\Len\calA_\Len{W}_\Len}\at\phase}-
\Phi\Bat{\at{\calA_\infty{}E_\Len{W}_\Len}\at\phase}
\,\dint{\phase}
\\&\leq%
\int\limits_{\babs{\abs\phase-\Len}\leq1/2}%
\Phi\Bat{\at{E_\Len\calA_\Len{W}_\Len}\at\phase}-%
\Phi\Bat{\at{\calA_\infty{}E_\Len{W}_\Len}\at\phase}
\,\dint{\phase}
\leq{f\at{\eps}},%
\end{align*}
with $f\at\eps=2\,\eps\sup\limits_{0\leq{r}
\leq\eps}\Phi^\prime\at{r}=\nDo{\eps}$. Now let
${W}_\infty\in\calS_{\infty,\,\ga}$ be fixed, and define
${W}_\Len\in\calS_{\Len,\,\ga}$ by
\begin{align*}
\at{{W}_\Len}\at\phase=
{W}_\infty\at\phase&\text{\;\;for $\abs{\phase}\le{\Len}$}.
\end{align*}
Then, in general we have
${W}_\infty\at\phase\neq{}\at{E_\Len{W}_\Len}\at\phase=0$ for
$\phase>\Len$ but always $E_\Len{W}_\Len\to{W}_\infty$ strongly in
$\fspaceL^2\at\Rset$ as $\Len\to\infty$. This implies
$\calP_\infty\at{E_\Len{W}_\Len}-\calP_\infty\at{{W}_\infty}\to0$,
and thanks to \eqref{GC:Lem.Conv1.Eqn0} we find
\eqref{GC:Lem.Conv1.Eqn1}.
\end{proof}
Lemma \ref{GC:Lem.Conv1} now provides the convergence of
suprema.
\begin{corollary}
\label{GC:Corr.Conv3}%
We have
\begin{math}
P_\Len\at{\gamma}
\xrightarrow{\Len\to\infty}P_\infty\at{\gamma}.
\end{math}
\end{corollary}
\begin{proof}
For given $\Len<\infty$ let $\ol{{W}}_\Len$ be a maximiser of
$\calP$ on $\calS_{\Len,\,\ga}$. Then \eqref{GC:Lem.Conv1.Eqn0}
implies
\begin{align*}
P_\Len\at{\ga}\leq%
\calP_\infty\at{E_\Len\ol{{W}}_\Len}+\nDo{\sqrt{\ga/\Len}}\,
\leq%
P_\infty\at\ga+\nDo{\sqrt{\ga/\Len}},
\end{align*}
and hence $\limsup_{\Len\to\infty}P_\Len\at\ga\leq{P}_\infty\at\ga$.
Moreover, in view of \eqref{GC:Lem.Conv1.Eqn1} we have
\begin{align*}
\calP_\infty\at{{W}_\infty}\leq\liminf_{\Len\to\infty}P_\Len\at\ga
\end{align*}
for all ${W}_\infty$, and this shows
$P_\infty\at{\gamma}\leq\liminf_{\Len\to\infty}P_\Len\at{\gamma}$.
\end{proof}
As a further consequence we find that solitons can be constructed as
limits of wave trains. More precisely, Corollary \ref{GC:Corr.Conv3}
combined with Corollary \ref{ES:Tightness.Corr2} provides the
following convergence result for maximisers.
\begin{corollary}
\label{GC:Corr.Conv4}%
Let $\calP$ be genuinely super-quadratic on $\calS_\gamma$, and for
each $\Len<\infty$ let $\ol{{W}}_{\Len}$ be a maximiser of $\calP$
in $\calS_{\Len,\,\ga}$. Then, for any sequence $\at{\Len_n}_n$ with
$\Len_n\to\infty$ there exist a subsequence, still denoted by
$\Len_n$, and a maximiser $\ol{{W}}_\infty\in\calS_{\infty,\,\ga}$,
such that $ E_{\Len_n}{\ol{{W}}_{\Len_n}}\to\ol{{W}}_\infty$
strongly in $\fspaceL^2\at{\Rset}$.
\end{corollary}
%
%
%
%
%
%
% ------------------------------------------------------------------
%                      acknowledgement
% ------------------------------------------------------------------
%
%
\subsection*{Acknowledgement}%
The author would like to thank Jens D.M. Rademacher for several stimulating discussions,
and Michael Helmers for providing the example of Remark
\ref{SQ:Rem.Convexity.Pot}.
%
%
% ------------------------------------------------------------------
%                      bibliography
% ------------------------------------------------------------------
%
\providecommand{\bysame}{\leavevmode\hbox to3em{\hrulefill}\thinspace}
\providecommand{\MR}{\relax\ifhmode\unskip\space\fi MR }
\providecommand{\MRhref}[2]{%
  \href{http://www.ams.org/mathscinet-getitem?mr=#1}{#2}
}
\providecommand{\href}[2]{#2}

%
% ------------------------------------------------------------------
%                     !!! end document !!!
% ------------------------------------------------------------------
%
\end{document}